# Zero Indirect Band Gap and Flat Bands in a Niobium Oxyiodide Cluster Material

Jan Beitlberger,[a] Mario Martin,[b] Marcus Scheele,[b] Marek Matas,[c] Carl P. Romao,*[c] Markus Ströbele,[a] and H.-Jürgen Meyer*[a]

[a] Section for Solid State and Theoretical Inorganic Chemistry, Institute of Inorganic Chemistry, Auf der Morgenstelle 18, 72076 Tübingen, Germany.

E-mail: juergen.meyer@uni-tuebingen.de

[b] Institute for Physical and Theoretical Chemistry, Eberhard Karls Universität Tübingen, Auf der Morgenstelle 18, 72076 Tübingen, Germany.

[c] Faculty of Nuclear Sciences and Physical Engineering, Czech Technical University in Prague, Czech Republic.

E-mail: carl.romao@cvut.cz

**ABSTRACT:** Explorative chemistry in a reaction system composed of $NbI_4$, $Li_2(CN_2)$, and $Li_2O$ has led to the discovery of a number of niobium oxyiodide cluster compounds. During this reaction, the formation of solid phases was detected alongside with gaseous phases, resulting in a range of products with cluster cores of varying shapes. After several niobium oxyiodide cluster compounds have already been identified within this reaction system, two additional compounds, $Nb_6O_3I_{15}$ and $Nb_{11}O_6I_{24}$, are discovered and structurally characterized by single-crystal X-ray diffraction. Both structures are based on the butterfly-shaped, oxygen-capped niobium cluster [$Nb_4O$], which is extended to larger cluster fragments. The [$Nb_4O$] cluster core in $Nb_6O_3I_{15}$ is extended by two [NbO] units to form a three-dimensional framework, and $Nb_{11}O_6I_{24}$ contains two connected [$Nb_4O$] units, which form chiral units within an antiferrochiral hexagonal packing of strings. The striking string-like character of $Nb_{11}O_6I_{24}$ was investigated in terms of its electronic structure and properties. DFT calculations showed $Nb_{11}O_6I_{24}$ to possess a zero indirect band gap, with a pair of 3-dimensional flat bands surrounding the Fermi level. These unusual features of the electronic band structure suggest the presence of strongly correlated inter-cluster singlet electron states, arising from the helical shape of the clusters, the hexagonal packing of the strings, and the delocalized nature of cluster electron wavefunctions.

## 1. INTRODUCTION

Transition metal clusters can exhibit a variety of structural arrangements in which the metal atoms can have multiple oxidation states. The arrangement of metal–metal bonds present in such clusters can give rise to many exotic electronic properties, which depend on the geometric arrangement of atoms within the clusters as well as that of the clusters relative to each other. Clusters can be used as optic,[1] superconducting,[2] electronic,[3] magnetic,[4] sensor,[1] and catalytic[5] materials due to the localization of their electrons and consequent strong electron correlation effects.[6]

One of the most prominent cluster types is the octahedral $M_6$ cluster, which commonly appears in [$M_6X_{12}$]- and [$M_6X_8$]-type architectures ($M$ = metal, $X$ = halide).[7] Therein, the octahedral metal cluster core is surrounded by twelve edge-capping $X$ atoms in [$M_6X_{12}$], and eight face-capping $X$ atoms in [$M_6X_8$]. As a result of the relative sizes of $M$ and $X$, the [$M_6X_{12}$]-type is obtained for a combination of large metal atoms and small halides and vice versa for the [$M_6X_8$]-type, with small metal atoms and large halides. [$M_6X_8$]-type clusters can form Chevrel phases (for $X$ = chalcogenides), which are well-known for their robust type-II superconductivity.[2]

Niobium halides include both cluster types, exemplified by $Nb_6Cl_{14}$ ([$M_6X_{12}$]) and $Nb_6I_{11}$ ([$M_6X_8$]).[8, 9] These compounds are typically prepared at comparatively high temperatures (700–950 °C), indicating their thermodynamic stability. A preferred synthesis route involves the metallothermic reduction of higher oxidized niobium halides with niobium metal. Variation of the oxidation state of these clusters can give rise to magnetic properties depending on the number of unpaired electrons.[10]

Binary niobium halides involve the compounds $Nb_3X_8$ ($X$ = Cl, Br, I)[11, 12], $NbX_3$ ($X$ = F, I)[13, 14], $NbX_4$ ($X$ = F, Cl, Br, I)[15-18] and $NbX_5$ ($X$ = F, Cl, Br, I)[19-22]. In all these compounds, the niobium centers are octahedrally coordinated by halide atoms; however, the overall structural connectivity depends strongly on the halide. For example, $NbF_5$ forms tetrameric units composed of corner-sharing [$NbF_4F_{2/2}$] octahedra[19], whereas the chloride and bromide analogues crystallize as dimeric, edge-sharing octahedral units[20, 21]. In niobium pentaiodide, either dimeric structures analogous to those of the chloride and bromide compounds are observed, or chains of corner-sharing [$NbI_4I_{2/2}$] octahedra are formed, arranged in a hexagonal packing arrangement.[22, 23]

Metal-rich niobium halide compounds, such as the Peierls-distorted $NbX_4$ ($X$ = Cl, Br, I) and $NbI_3$, exhibit string-like connectivities.[14, 16-18] In contrast, the $Nb_3X_8$ compounds feature layered structures based on triangular $Nb_3$ clusters.[11, 12] The electronic properties of the $Nb_3X_8$ compounds have been extensively studied due to the presence of flat electronic bands in single-layer $Nb_3Cl_8$, and a complex interplay between Mott insulating behavior and singlet formation in the bulk.[4, 24, 25]

The introduction of another anion, alongside the halide, leads into the field of heteroanionic compounds, which significantly expands the structural and chemical diversity of these compounds. In many instances, the heteroanion substitutes for a halide within an existing framework, thereby modifying the connectivity within the crystal structure. Examples include $Nb_6SI_9$[26], derived from $Nb_6I_{11}$, as well as $Nb_3SX_7$ ($X$ = Cl, Br, I)[27-29] and $ANb_3SBr_7$ ($A$ = Rb, Cs)[30, 31] whose central clusters can be derived from $Nb_3Br_8$[12]. The one-dimensional chains of clusters in $ANb_3SBr_7$ lead to the formation of an electronic Luttinger liquid, a type of quantum metal.[31]

Alternatively, the incorporation of heteroatoms can lead to the formation of clusters with distinct shapes. Especially $Nb_4$ clusters have shown to adopt various shapes such as square, rectangular, rhombohedral, tetrahedral, and butterfly-type cluster core geometries.[32-36] The introduction of 4*f* heteroatoms to 3*d* organometallic butterfly clusters has been used to synthesize many single-molecule magnets.[37]

Butterfly-shaped metal clusters can be related to an octahedral $[M_6X_{12}]$-type cluster with two missing metal edges. This type of cluster is usually ($\mu_4$-) capped by a heteroatom, as can also be derived from the $[M_6ZX_{12}]$-type cluster[38-42], possessing an interstitial $Z$ heteroatom, by removing two metal corners. Butterfly clusters of this type are well-established; most of them were prepared by means of solution chemistry.

A first example was reported by Manassero *et al.*, who characterized the compound $[Me_3NCH_2Ph][Fe_4(CO)_{13}H]$ (Me = methyl group; Ph = phenyl group).[43] It features a $Fe_4$ core arranged in a butterfly geometry, capped by a carbonyl ligand. The cluster possesses a total number of 12 electrons available for metal–metal bonding. Further synthetic developments have expanded the chemistry of these clusters. For instance, oxidative reactions of the carbon-centered cluster $[Fe_6C(CO)_{16}]^{2-}$ have yielded compounds such as $(Et_4N)[Fe_4C(CO)_{12}\cdot CO_2CH_3]$[44] (Et = ethyl group) and $Fe_4C(CO)_{13}$[45].

The butterfly motif is not limited to iron-based systems; analogous clusters have been identified for various transition metals. One notable example is $W_4C(O^iPr)_{12}(NMe)$ ($^iPr$ = isopropyl group), a tungsten cluster that contains only six electrons for metal–metal bonding.[46] This low electron count results in relatively long W–W distances, averaging at 2.78 Å. In contrast, the molybdenum cluster $Mo_4Br_4(O^iPr)_8$ adopts a related butterfly structure, but with 12 electrons for Mo–Mo bonding, leading to significantly shorter average Mo–Mo bond lengths of approximately 2.50 Å.[47]

Compounds containing butterfly clusters were also obtained by means of solid-state reactions under moderate heating conditions (as low as 400 °C). The series of compounds $Nb_4PnX_{11}$ ($Pn$ = N, P; $X$ = Cl, Br, I)[36] features $\mu_4$-pnictogen-capped butterfly cores, where the central pnictide atom bridges all four niobium atoms. A $Ta_4$ based butterfly cluster with a $\mu_4$-capping sulfur atom has been observed in the tantalum compound $Ta_4SBr_{11}$; magnetic correlations between these clusters create a Mott insulating state.[48]

Flexibility in the number of electrons available for metal–metal bonding interactions is widespread in cluster compounds, especially for niobium. An example is the octahedral $[Nb_6Cl_{18}]^{x-}$ cluster ($x$ = 2, 3, 4), where the successive reduction of cluster electrons leads to an elongation of the Nb–Nb bonds.[49] Another cluster system with variable electron counting is reported for triangular $Nb_3$ clusters. The valence electron concentration (VEC) varies, with 6 ($Nb_3SX_7$ with $X$ = Cl, Br, I;[27-29] $(PEt_3H)[Nb_3Cl_{10}(PEt_3)_3]$[50]), 7 ($Nb_3X_8$ with $X$ = Cl, Br, I;[11, 12] $ANb_3SBr_7$ with $A$ = Rb, Cs[30, 31]), 7.5 ($AVNb_3Cl_{11}$ with $A$ = K, Rb, Cs, Tl[51]), and 8 cluster electrons ($Nb_3Cl_7(PMe_2Ph)_6$[50] and $NaNb_3Cl_8$[52]) having been reported. Also, a variable halide content can lead to different numbers of cluster electrons, as demonstrated by the $Nb_4OI_{12-x}$ series, which includes the members $Nb_4OI_{12}$ (VEC = 6), a- and b-$Nb_4OI_{11}$ (VEC = 7), and $Nb_4OI_{10}$ (VEC = 8).[53]

Heteroanionic cluster compounds are generally more thermally labile than their octahedral counterparts based on $[M_6X_{12}]$ and $[M_6X_8]$ clusters. Therefore, their synthesis requires carefully controlled reduction conditions, using a suitable reduction agent and reaction temperatures. Recently, a series of niobium oxyiodide clusters was obtained by reducing $NbI_4$ with $Li_2(CN_2)$, in the presence of $Li_2O$ as an oxide source under closely related temperature conditions near 500 °C, yielding compounds like $Nb_4OI_{12}$, $Nb_8O_5I_{17}(NbI_5)$, and others.[54-56]

These compounds appear to be metastable products, which decompose at elevated temperatures. We report the synthesis and characterization of two more compounds in this system, $Nb_6O_3I_{15}$ and $Nb_{11}O_6I_{24}$. These compounds contain a previously unreported structural feature: oxygen-capped $[NbO_4]$ butterfly cluster units, which, in conjunction with [NbO], form extended, asymmetric clusters. Analysis of the electronic structure and properties of $Nb_{11}O_6I_{24}$ reveals that the cluster shape and packing lead to flat electronic bands, which are perturbed to create an indirect band gap with zero magnitude.

## 2. Results and Discussion

### 2.1 Synthesis and Crystal Structure

The reaction between $NbI_4$, $Li_2(CN_2)$ and $Li_2O$ involves both solid and gaseous phases, and is governed by temperature-dependent equilibria. On heating, the mixture generates a sequence of intermediate solid and vapor phases. The collective experimental observations have led to an assumption of a process that is close to a non-equilibrium system.[57] The product formation is highly sensitive to the reaction temperature, duration, and heating and cooling rates, which determine the temporal exposure of reactants and intermediates to local thermal and chemical environments. Slight changes in these parameters lead to the crystallization of different niobium oxyiodide clusters with the rectangular $[Nb_4O]$, $[Nb_5O_4]$, and $[Nb_8O_5]$ cores known from earlier work[53-56], to the newly discovered butterfly-based $[Nb_6O_3]$ and $[Nb_{10}O_4]$ architectures. Such temperature-dependent selectivity indicates kinetic control of the product formation, with multiple competing reaction pathways accessible within a narrow thermal window.

Moreover, several products are metastable, decomposing upon extended heating or exposure to higher temperatures. Such metastable phases can only be stabilized under specific kinetic conditions or by rapid cooling. The persistence of these phases is governed not by thermodynamic stability, but by kinetic barriers that hinder transformation into more stable forms.

Thereby, the product formation is a result of competing kinetic processes, occurring under a continuous interplay between solid-state reactions and gas-phase transport.

The newly isolated compounds $Nb_6O_3I_{15}$ and $Nb_{11}O_6I_{24}$ require sensitive reaction conditions, close to those obtained for $Nb_8O_5I_{17}(NbI_5)$.[56] Both compounds were obtained from a reaction of $NbI_4$ with $Li_2O$ and $Li_2(CN_2)$ in 2:1:1 molar ratio. The reaction mixture was heated to 500 °C for one hour, followed by a controlled cooling to 450 °C with a rate of 1 °C/min, and then further cooled to room temperature with a rate of 0.1 °C/min. Both compounds crystallize as black solids. $Nb_6O_3I_{15}$ forms block-shaped crystals (Figure S1, top), while $Nb_{11}O_6I_{24}$ crystallizes as elongated platelets (Figure S1, bottom).

The structures of both compounds were determined by single-crystal X-ray diffraction, with corresponding crystallographic data and refinement parameters summarized in Table 1.

**Table 1. Crystallographic data from X-ray single-crystal refinement on $Nb_{18}O_9I_{45}$ and $Nb_{11}O_6I_{24}$.**

| | $Nb_6O_3I_{15}$ | $Nb_{11}O_6I_{24}$ |
|---|---|---|
| CCDC No. | 2401147 | 2380623 |
| sum formula | $Nb_{18}O_9I_{45}$ | $Nb_{11}O_6I_{24}$ |
| Space group | $C2/c$ | $P2_1/c$ |
| Temperature (K) | 150 | 150 |
| Unit cell dimensions (Å) | $a$ = 26.9643(4) | $a$ = 16.4086(1) |
| | $b$ = 14.3399(2) | $b$ = 16.2749(2) |
| | $c$ = 24.8587(4) | $c$ = 18.8946(2) |
| Monoclinic angle (°) | $\beta$ = 95.744(1) | $\beta$ = 91.713(1) |
| Volume (Å$^3$) | 9563.7(2) | 5043.52(9) |
| Z | 4 (Z' = 12) | 4 |
| Wavelength (Å) | 0.71073 | 1.54184 |
| $\mu$ (mm$^{-1}$) | 16.616 | 135.138 |
| Calculated density (g/cm$^3$) | 5.228 | 5.483 |
| 2$\theta$ range for data collection | 4.25 to 60.062 | 5.388 to 127.596 |
| Total number of reflections | 58082 | 44314 |
| Independent reflections | 13984 | 8139 |
| Refined parameters | 431 | 370 |
| $R_{int}$ | 0.0192 | 0.0236 |
| $R_1$ (I≥2$\sigma$(I) / all) | 0.0236 / 0.0286 | 0.0292 |
| $wR_2$ (I≥2$\sigma$(I) / all) | 0.0551 / 0.0565 | 0.0745 |
| Goodness-of-fit on $F^2$ | 1.031 | 1.020 |

### 2.1.1 The Crystal Structure of $Nb_6O_3I_{15}$

The compound $Nb_6O_3I_{15}$ features the presence of two slightly different [$Nb_4O$] cluster cores, which can be described as oxygen-capped butterfly clusters. Both butterfly clusters are extended by two [NbO] units each, to yield [$Nb_4O(NbO)_2$] fragments. The shapes of these two clusters are shown in Figure 1 and interatomic distances are collected in Table 2.

The Nb–Nb backbone, connecting the wings of the butterfly (Nb2–Nb5 and Nb8–Nb8'), is slightly longer (2.9552(7) Å and 2.968(1) Å, respectively) than the other Nb–Nb bonds (2.8850(7)–2.9184(7) Å) in the structure. A related cluster can be found in the structures of $Nb_4PnX_{11}$ ($Pn$ = N, P; $X$ = Cl, Br, I)[36], where the butterfly clusters are capped by a pnictide atom, as [$Nb_4Pn$]. Here, a contrary situation appears, as the backbone linkage is the shortest Nb–Nb bond. For example, in $Nb_4NBr_{11}$, the Nb–Nb backbone bond (2.928(3) Å) is slightly shorter than the other Nb–Nb distances (3.003(2) Å). A similar situation can be found in the $\mu_4$-S-capped $Ta_4$ cluster in the structure of $Ta_4SBr_{11}$.[48] A direct comparison of these clusters is, however, limited, as the connectivity pattern varies and their valence electron counts, from a classical point of view, differ with VEC = 9 for $Nb_6O_3I_{15}$, VEC = 7 for $Ta_4SBr_{11}$, and VEC = 6 for $Nb_4PnX_{11}$.

The Nb–Nb distances in both crystallographically independent [$Nb_4O$] butterfly clusters of $Nb_6O_3I_{15}$ are almost the same (Table 2). The Nb–Nb distances with the additional [NbO] extensions are somewhat longer and are showing some surprising differences when comparing both [$Nb_4O(NbO)_2$] cluster cores. The distances Nb1–Nb2 and Nb5–Nb6 are nearly the same, but notably shorter than the Nb8–Nb9 contacts in the parent cluster. This disparity could reflect differences in the electronic distribution, assuming that the cluster exhibiting the shorter Nb–Nb contacts to the [NbO] units possesses higher electron density than the one with the longer connection. An intramolecular charge separation has recently been discussed for the low-temperature polymorph of the van der Waals layered compound $Nb_3Cl_8$, exhibiting alternating layers of $[Nb_3]^{7+}$ and $[Nb_3]^{9+}$ clusters.[58] Another explanation could be attributed to matrix effects, influencing the Nb–Nb distances.

A butterfly-type niobium cluster with a capping oxygen atom has not been reported previously. The closest known analogue is found in $Nb_4OTe_9I_4$, which contains a flattened, oxygen-centered tetrahedral $Nb_4$ cluster.[59] Due to the availability of only four cluster electrons for Nb–Nb bonding, the Nb–Nb distances therein are relatively long (3.050(3)–3.057(4) Å).

The overall crystal structure of $Nb_6O_3I_{15}$ features a [$Nb_4O(NbO)_2$] cluster core that is interconnected with four neighboring clusters via pairs of $\mu_2$-iodide bridges (Figure 2). Two of these bridging connectivities occur at the "wingtips" of the butterfly cluster, connecting along the $a$- and $b$-axis directions, and two are at the [NbO] elongation, connecting along the $c$-axis. This connectivity of clusters results in the formation of a three-dimensional framework, displayed in Figure 3.

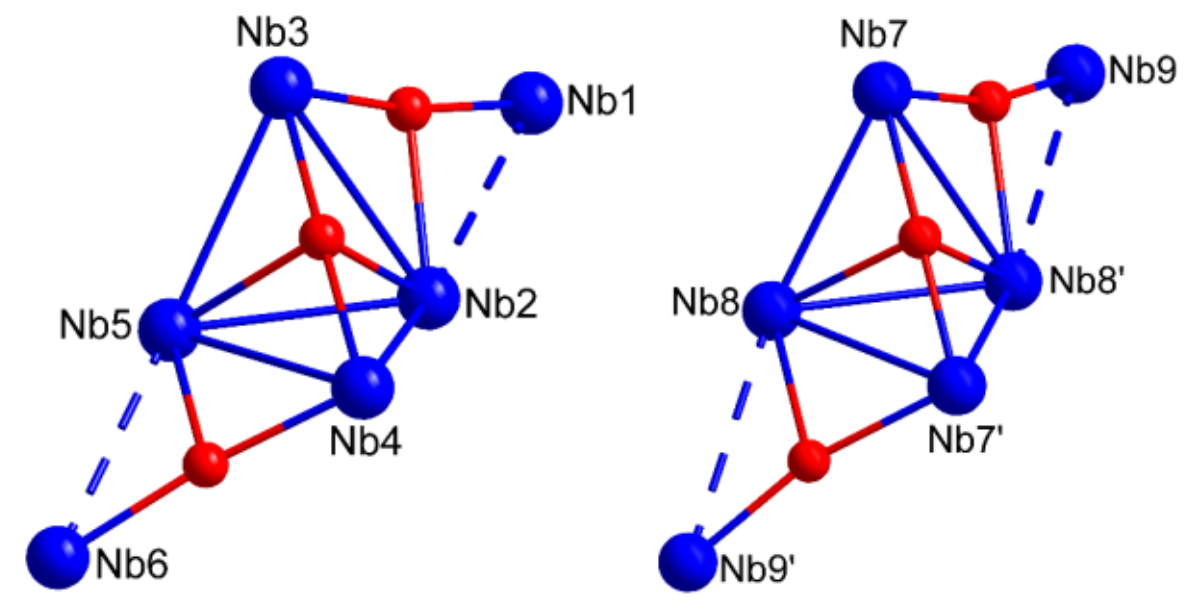

**Figure 1.** Two crystallographically distinct [$Nb_4O$] clusters extended by two [NbO] units in the structure of $Nb_6O_3I_{15}$.

**Table 2. Comparison of corresponding Nb–Nb distances of two distinct [$Nb_4O(NbO)_2$] cluster cores of $Nb_6O_3I_{15}$ compared in the left and right column.**

| Atoms | Distance/Å | Atoms | Distance/Å |
| --- | --- | --- | --- |
| Nb3–Nb5 | 2.9184(7) | Nb7–Nb8 | 2.8867(7) |
| Nb3–Nb2 | 2.9081(7) | Nb7–Nb8’ | 2.8880(7) |
| Nb2–Nb4 | 2.8850(7) | Nb7’–Nb8’ | 2.8867(7) |
| Nb4–Nb5 | 2.8945(7) | Nb7’–Nb8 | 2.8880(7) |
| Nb2–Nb5 | 2.9552(7) | Nb8–Nb8’ | 2.968(1) |
| Extensions of [$Nb_4O$] fragments | | | |
| Nb1–Nb2 | 3.1017(7) | Nb8’–Nb9 | 3.2720(7) |
| Nb5–Nb6 | 3.1027(7) | Nb8–Nb9 | 3.2720(7) |

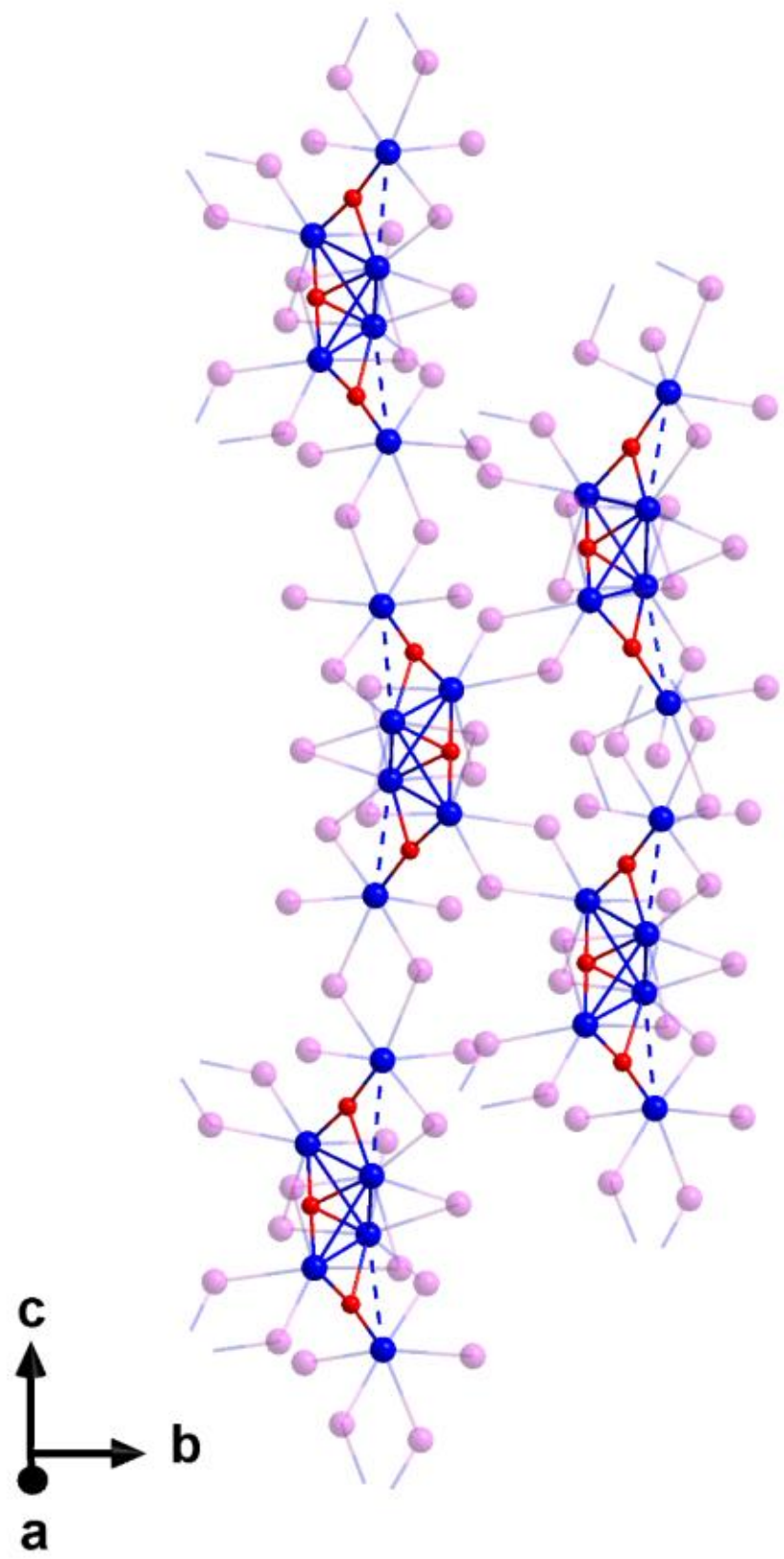

**Figure 2.** Connectivity of the [$Nb_4O(NbO)_2$] cluster in the structure of $Nb_6O_3I_{15}$ with $\mu_2$-iodide bridges. Niobium atoms are colored in blue, oxygen in red and iodine in light pink.

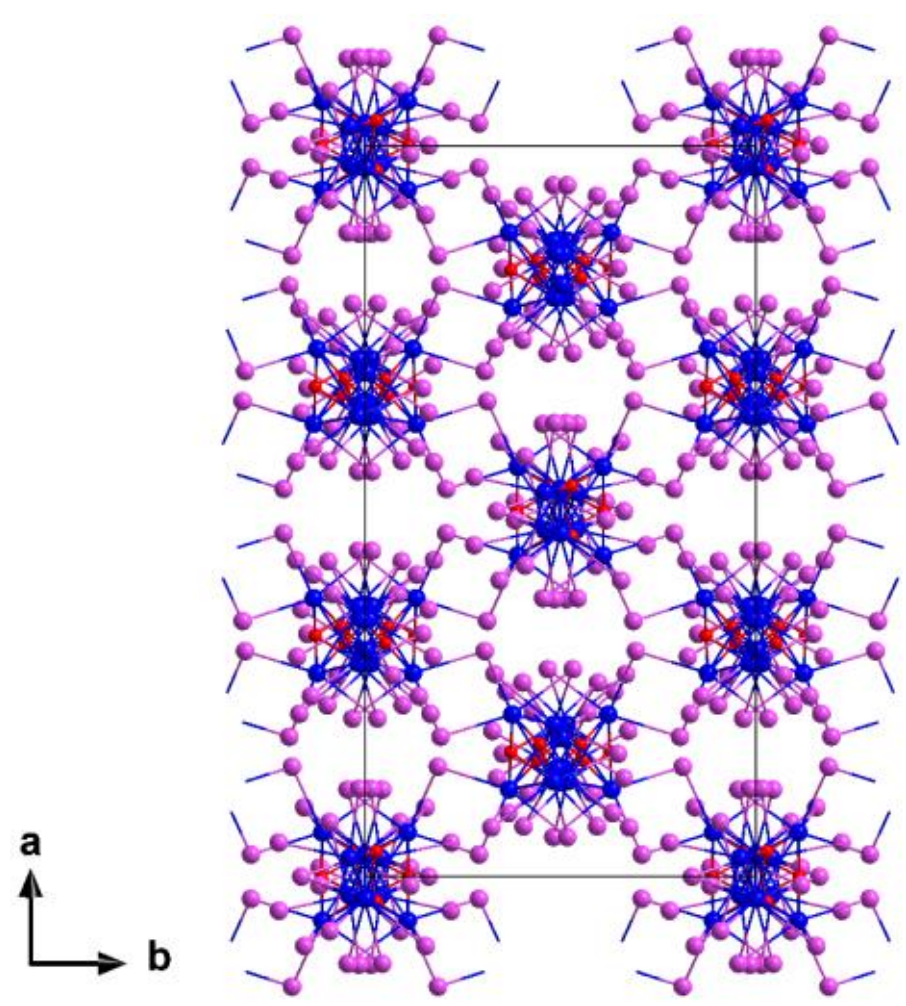

**Figure 3.** Section of the crystal structure of $Nb_6O_3I_{15}$, showing the formation of a three-dimensional network.

### 2.1.2 The Crystal Structure of $Nb_{11}O_6I_{24}$

The crystal structure of $Nb_{11}O_6I_{24}$ is based on a [$Nb_{10}O_4$] core built from two independent butterfly [$Nb_4O$] clusters of the same type as the previous structure of $Nb_6O_3I_{15}$. A comparison of the extended butterfly cluster [$Nb_4O(NbO)_2$] in $Nb_6O_3I_{15}$ with the [$Nb_{10}O_4$] core in $Nb_{11}O_6I_{24}$ is shown in Figure 4. Two [$Nb_4O$] clusters in the structure of $Nb_{11}O_6I_{24}$ are interconnected by a pair of niobium atoms (Nb6 and Nb7) as ($ONb_2O$) to form the [($Nb_4O$)($ONb_2O$)($Nb_4O$)] fragment, which is interconnected with two adjacent clusters by one niobium atom (Nb1) as [ONbO] to yield the infinite chain structure $^1_\infty$[($Nb_4O$)($ONb_2O$)($Nb_4O$)(ONbO)], displayed in Figure 5. The two [$Nb_4O$] units are mirrored relative to each other, leading to a twist of the overall [$(Nb_4O)_2(Nb_2O_2)$] unit, giving it a helical shape. An analysis of interatomic Nb–Nb distances is of particular interest, as they reflect the electronic interaction of 19 cluster electrons within and between butterfly clusters of $Nb_{11}O_6I_{24}$. These distances show notable variations, as summarized in Table 3.

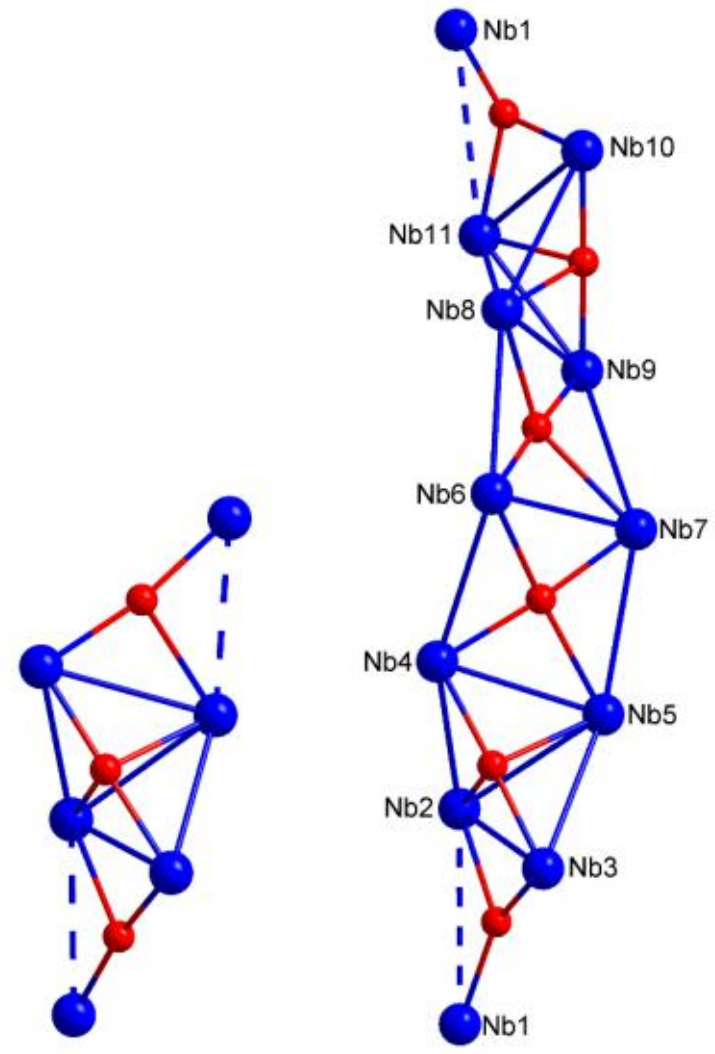

**Figure 4.** Comparison of cluster fragments in structures of $Nb_6O_3I_{15}$ (left) and $Nb_{11}O_6I_{24}$ (right). Corresponding Nb–Nb bond lengths of $Nb_{11}O_6I_{24}$ are summarized in Table 3.

**Table 3. Nb–Nb bond distances inside the $[(Nb_4O)_2(Nb_2O_2)]$ core of $Nb_{11}O_6I_{24}$.**

| Atoms | Distance/Å | Atoms | Distance/Å |
|---|---|---|---|
| $[Nb_4O]$-Fragments | | | |
| Nb2–Nb4 | 2.7941(9) | Nb8–Nb10 | 3.0032(9) |
| Nb4–Nb5 | 2.8632(9) | Nb10–Nb11 | 2.8995(9) |
| Nb3–Nb5 | 3.0128(9) | Nb9–Nb11 | 2.7913(9) |
| Nb2–Nb3 | 2.9176(9) | Nb8–Nb9 | 2.8697(9) |
| Nb2–Nb5 | 2.9951(9) | Nb8–Nb11 | 2.9961(9) |
| Interconnection between $[Nb_4O]$-fragments | | | |
| Nb4–Nb6 | 3.0759(9) | Nb6–Nb8 | 3.082(1) |
| Nb5–Nb7 | 3.071(1) | Nb7–Nb9 | 3.072(1) |
| Nb6–Nb7 | 2.9230(9) | | |
| Interconnection of $[Nb_{10}O_4]$ clusters | | | |
| Nb1–Nb2 | 3.5110(9) | Nb1–Nb11 | 3.4230(9) |

The Nb–Nb distances of the $[Nb_4O]$-fragments (2.7913(9)–3.0128(9) Å) are showing a somewhat wider distribution of distances than those in $Nb_6O_3I_{15}$ (2.8850(7)–2.968(1) Å, see Table 2). This may be caused by the interconnection of the butterfly-clusters via Nb6/Nb7 atoms within the $[Nb_{10}O_4]$ fragments, with Nb–Nb contacts ranging between 3.071(1)–3.082(1) Å. The $[Nb_{10}O_4]$ fragment is interconnected with two adjacent clusters through another niobium atom (Nb1) at longer distances (Nb1–Nb2 = 3.5110(9) Å and Nb1–Nb11 = 3.4230(9) Å) shown in Figure 5.

The overall crystal structure of $Nb_{11}O_6I_{24}$ is composed of cluster strings running parallel to the *b*-axis, following a hexagonal stick packing motif (Figure 6). Each single string is separated by a van der Waals gap. This type of one-dimensional chain structure is also found in other cluster based compounds, such as $Nb_6I_9S$[26] whose structure consists of sulfur-bridged $[M_6X_8]$-type octahedra, as well as a- and b-$Nb_4OI_{11}$[53], where rectangular $[Nb_4O]$ clusters are interconnected into a one-dimensional chain structure. These anion-bridged compounds show semiconducting behavior with small band gaps. In contrast, the one-dimensional chains in the structure of $A$Nb$_3$Br$_7$S ($A$ = Cs, Rb) are separated by alkali ions and the material exhibits semimetallic behaviour with characteristics of a Luttinger liquid.[31]

To gain insights into the electronic structure of $Nb_{11}O_6I_{24}$, its electrical conductivity and the electronic band structure were investigated.

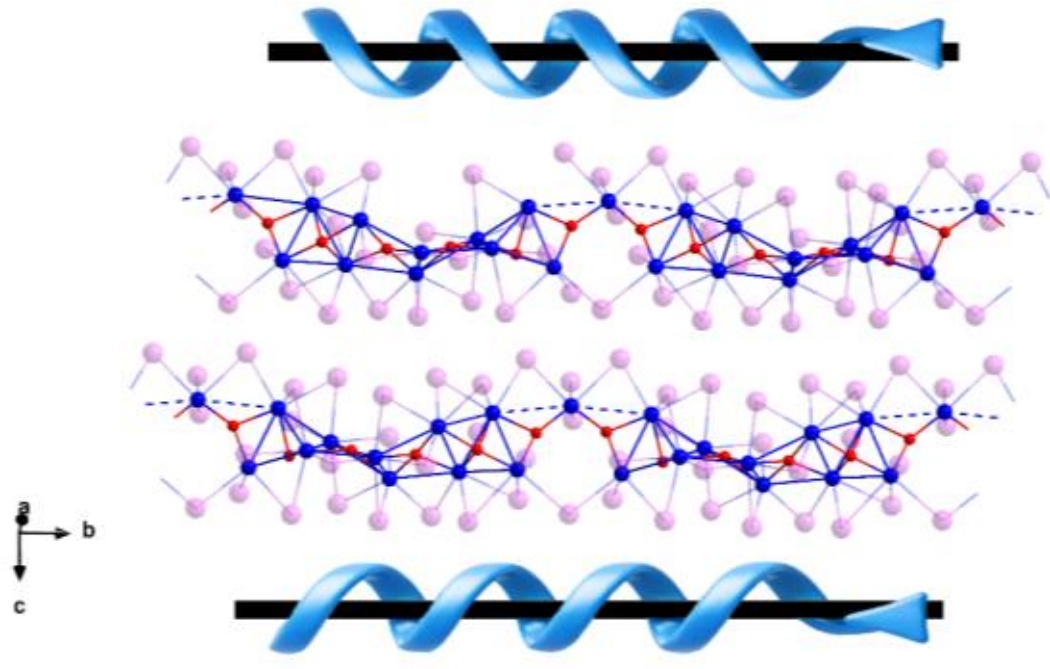

**Figure 5.** Helical-shaped and string-like connectivity of the clusters in $Nb_{11}O_6I_{24}$. Niobium atoms are colored in blue, oxygen in red and iodine in light pink.

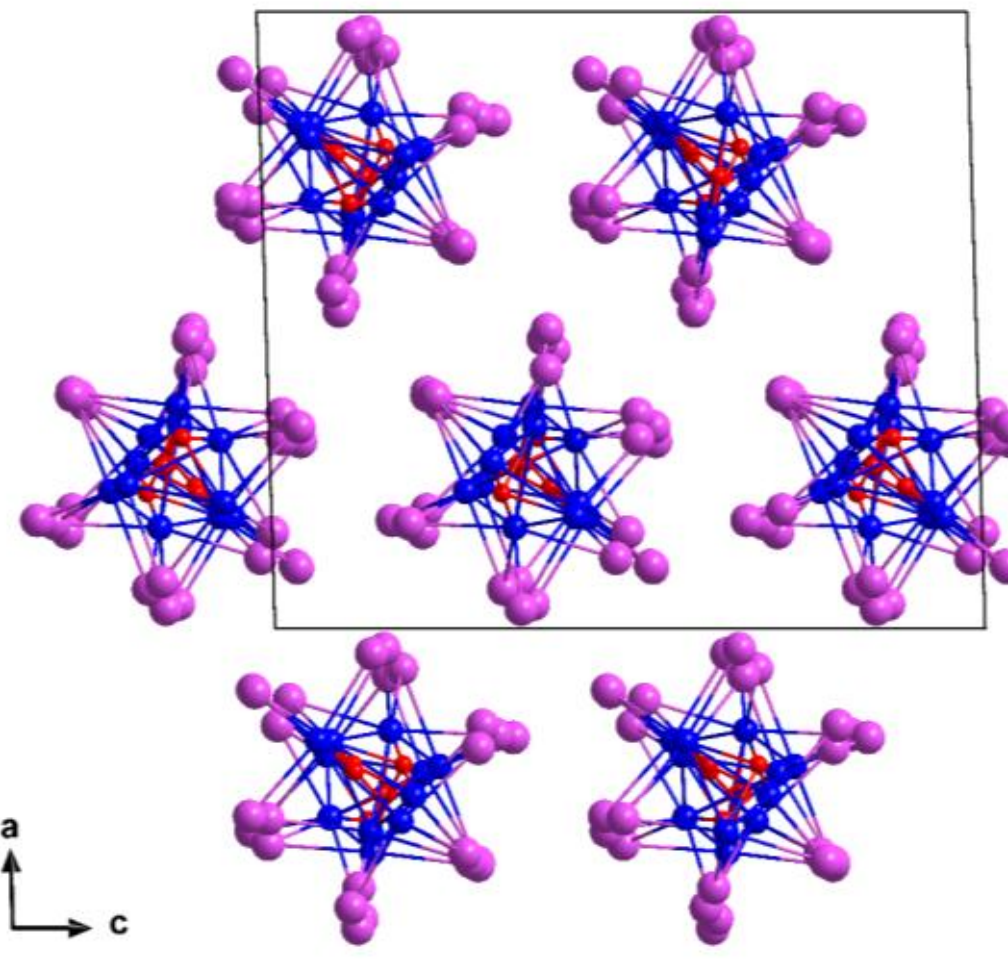

**Figure 6.** Section of the crystal structure of $Nb_{11}O_6I_{24}$ along the *b*-axis, illustrating the hexagonal packing of the strings.

## 2.2 Electronic Structure

The electronic band structures of $Nb_6O_3I_{15}$ (Figure 7) and $Nb_{11}O_6I_{24}$ (Figure 8) were calculated using density functional theory (DFT). In both cases, the bands near the Fermi energy correspond to Nb 4*d* orbitals; I 5*p* orbitals appear at lower energies (see Supporting information, Figures S1–S3). $Nb_6O_3I_{15}$ is found to be a metal, whereas $Nb_{11}O_6I_{24}$ has an indirect zero band gap between the valence band maximum (VBM) at D (0 ½ ½) and the conduction band minimum (CBM) at E (−½ ½ ½).

This finding of an indirect zero band gap is highly unusual. Many zero-gap materials are known, including doped semiconductors, Dirac materials, and topological insulators; however, in these experimentally realized materials, the zero gap is direct.[60] The study of systems with indirect zero band gaps has largely been limited to theoretical proposals based on model Hamiltonians,[61-66] with the exception of a recent experimental realization in a photonic metamaterial.[67] Therefore, the DFT calculations suggest that $Nb_{11}O_6I_{24}$ holds a unique status as an atomic crystal with an inherent zero indirect band gap.

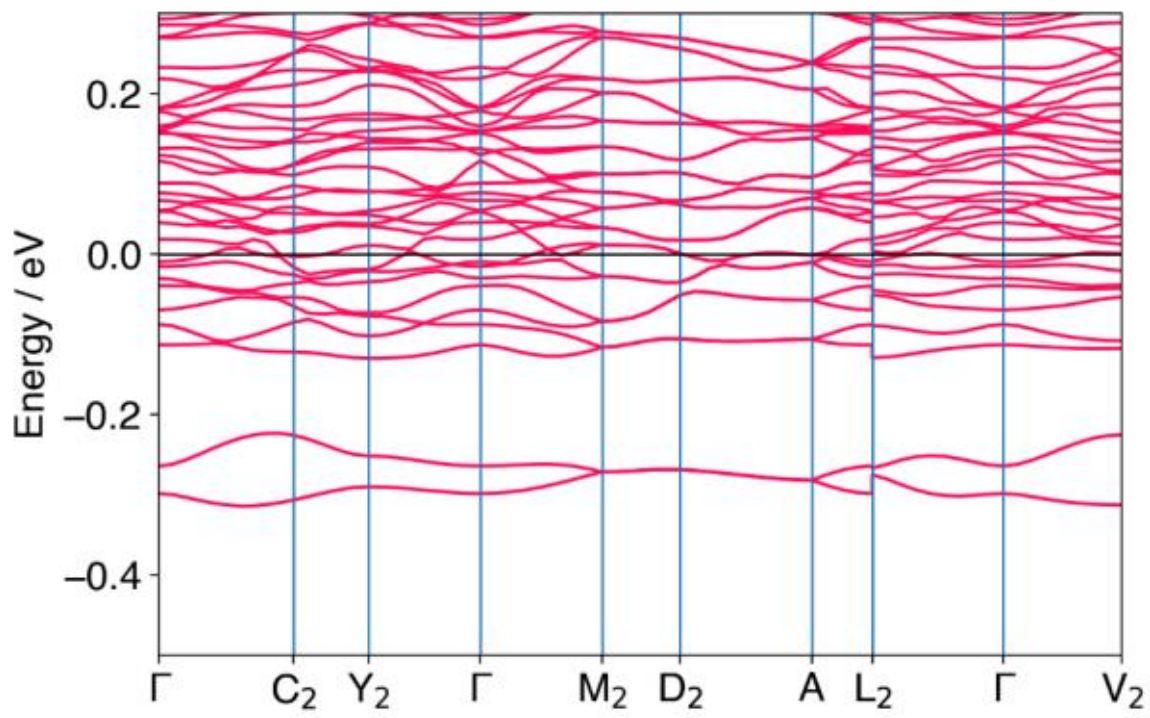

**Figure 7.** The electronic band structure of $Nb_6O_3I_{15}$, showing metallic bands formed from Nb 4*d* orbitals. Special points in and paths through the Brillouin zone were chosen following the literature.[68]

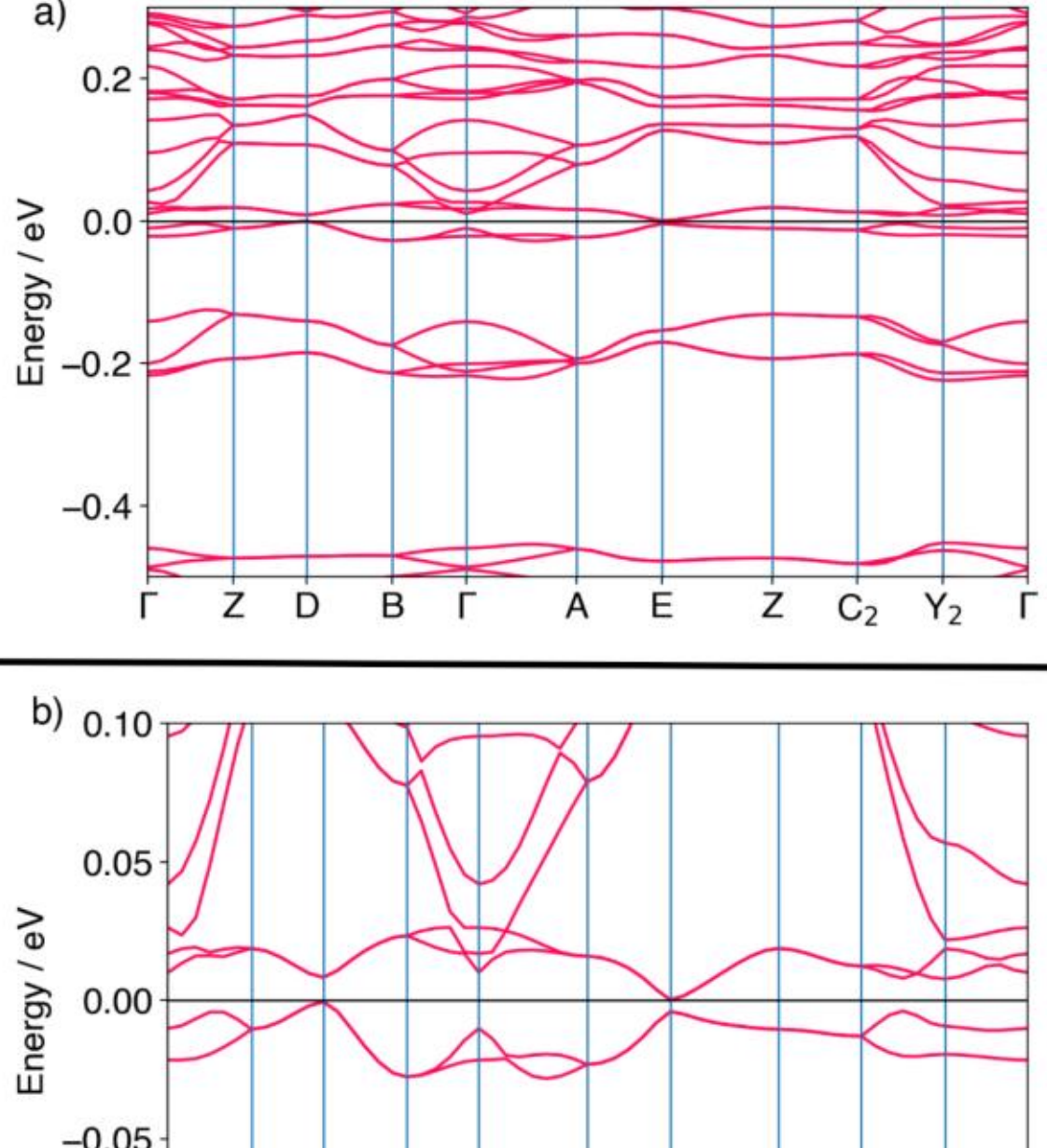

**Figure 8.** The electronic band structure of $Nb_{11}O_6I_{24}$, showing the band manifolds near the Fermi energy (a), and a closeup of the flat bands surrounding the Fermi energy and the indirect zero gap between D (0 ½ ½) and E (−½ ½ ½) (b). Special points in and paths through the Brillouin zone were chosen following the literature.[68]

Additionally, the electronic bands near the Fermi energy are nearly flat, corresponding to spatially localized wavefunctions. Such localized states can enhance electronic correlation effects, which have been associated with phenomena such as unconventional superconductivity[69] and the fractional quantum Hall effect.[70] Flat bands have been studied extensively in the 2-dimensional cluster compound $Nb_3Cl_8$[4, 24] and related materials.[25, 71] The strongly correlated flat bands in $Nb_3Cl_8$ do not lead to superconductivity, which allows the material to be used as a field-free Josephson diode.[72, 73]

The effective dimensionality of the electronic structure of $Nb_{11}O_6I_{24}$ is therefore of interest, because zero-dimensional flat bands, corresponding to the atomic limit, would fail to produce the quantum effects commonly associated with flatbands.[74] The bands near the Fermi energy are flat in all directions in reciprocal space (Figure 8), indicating either zero or three-dimensional character. However, since each $Nb_{11}O_6I_{24}$ structural unit possesses an odd number of electrons, the absence of a metallic state is a clear indication that the clusters do not behave like isolated atoms. The formation of a singlet state indicates inter-cluster interactions along the *a* and *c* directions, since each unit cell contains only one repetition of the $Nb_{11}O_6I_{24}$ unit along each string along *b*.

The dimensionality and localization of the wavefunction can be seen through the maximally localized Wannier functions (MLWFs),[75] which are shown for the valence band maximum in Figure 9a,c and for the conduction band minimum in Figure 9b,d, based on Wannierization of the band manifold between −0.05 and 0.35 eV (see Supporting Information, Fig. S6). The MLWFs show that the electronic states are delocalized over the $[(Nb_4O)_2(Nb_2O_2)]$ clusters and into the interstitial space, with some reach into neighboring clusters. The delocalization thereby allows adjacent clusters to interact, forming a quantum singlet state.

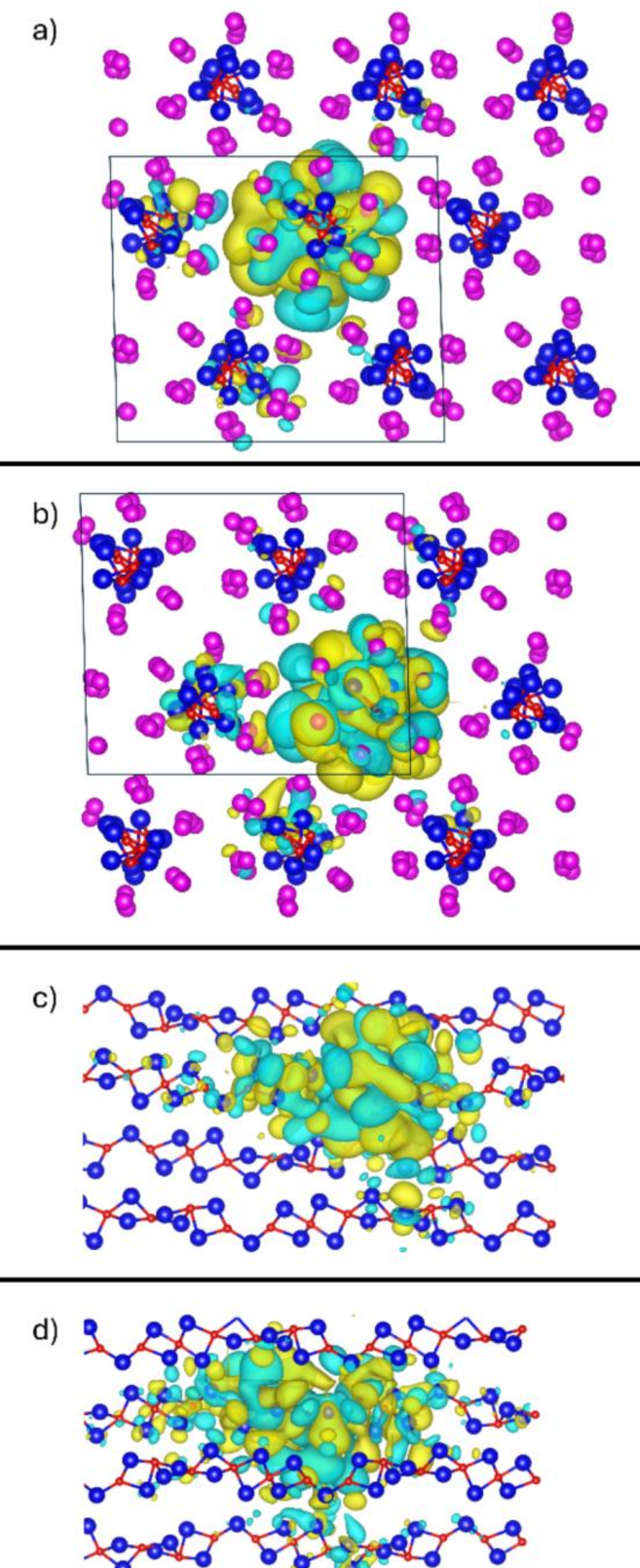

**Figure 9.** Maximally localized Wannier functions of the highest occupied (a, c) and lowest unoccupied (b, d) bands in $Nb_{11}O_6I_{24}$, shown along *b* (a, b) and along *a* (c, d). Iodide atoms are omitted from (c, d) for clarity. The MLWFs are delocalized over $[(Nb_4O)_2(Nb_2O_2)]$ clusters and into the interstitial space, allowing electron–electron interactions that lead to the formation of a singlet state.

The nature of the states near the Fermi energy, as well as the stability of the singlet state, were analysed further through calculation of the crystal orbital Hamilton populations (COHP).[76, 77] The COHP decomposes the electronic states into bonding and antibonding orbitals, showing the states near the Fermi level to be comprised of Nb–O and Nb–I antibonding orbitals and Nb–Nb bonding orbitals, a situation which persists up to about −2 eV, where Nb–I bonding orbitals become predominant (Figure S7). At the Fermi energy, the COHP of the bonding and antibonding orbitals are essentially balanced, indicating that the quantum singlet state is energetically stable at the DFT level. The covalent bond strengths (as represented by the integrated COHP) of the Nb–O, Nb–I, and Nb–Nb bonds all ranged from −2 to –3 eV, showing that all three bond types play a role in the formation of the complex $[(Nb_4O)_2(Nb_2O_2)]$ cluster structure.

We have also investigated the topology of the flat bands as another potential origin of the zero indirect gap.[61, 62] Topological flat bands can show the fractional quantum Hall effect and quantum spin Hall effect, in addition to having topological protection of the band gap.[78-80] Calculation of the $\mathbb{Z}_2$ topological invariant using the Wilson loop method[81] revealed nontrivial topology in the $k_a k_c$ plane, corresponding to a weak topological insulator with $\mathbb{Z}_2$ = (0; 010).[82] This finding indicates that $Nb_{11}O_6I_{24}$ behaves like a series of stacked 2D topological insulators,[80, 83] each of which has the hexagonal structural motif typical of 2D topological flatband materials.[74, 84]

At this point, we can connect the electronic properties of $Nb_{11}O_6I_{24}$ to its structure. Once we have established the three-dimensional nature of the electronic structure, the destructive interference of the wavefunction, which creates flat bands in the *ac* plane, can be assumed to arise from the hexagonal structural motif, as is seen in Kagome materials and twisted bilayer graphene.[74, 84] In the vicinity of D and E in Figure 8, we find the flat band picture to be perturbed, creating the indirect zero band gap. The origin of this unusual gap can be related to the helical shape of the clusters (Figure 4), as chirality and helicity are known to play an important role in the formation of such gaps in model systems and photonic crystals.[61, 66, 67]

Specifically, in these model systems local symmetry breaking appears in the form of terms in the Hamiltonian which move the VBM and CBM to different points in momentum space.[61, 66,67] Meanwhile, the retention of global symmetry prevents the opening of a topologically trivial gap and thereby the system acquires a zero indirect gap.[61, 66, 67] We can identify these features in $Nb_{11}O_6I_{24}$, which contains two pairs of chiral clusters with opposite helicity within each unit cell, leading to an overall antiferrochiral[85] arrangement which retains global inversion symmetry. In contrast, other Nb cluster compounds which lack the helical motif do not show indirect band gaps, including examples such as $Nb_4OI_{10}$ and $CsNb_3Br_7S$, which also possess flat bands.[31, 54-56] $Nb_{11}O_6I_{24}$ is distinguished from the model systems by the preservation of time reversal symmetry (TRS); in the models TRS is broken, creating a zero indirect gap between TRS-related points at $\pm\mathbf{k}$.[61, 66, 67]

Inclusion of spin–orbit coupling (SOC) in the DFT calculations causes only minor changes to the bands near the Fermi level (Figure S8), as the spin–orbit interaction is weak in the Nb 4*d* states. However, these changes are nevertheless instructive regarding the origins of the indirect zero band gap. With SOC, the band gap takes a very small negative value (*ca.* −3 meV). Such band inversion due to SOC is well-known in direct zero gap materials, where it can cause the opening of a positive gap, because crossing of the inverted bands becomes forbidden.[86, 87] In $Nb_{11}O_6I_{24}$, band inversion can create a negative gap as the bands are separated in momentum space. Band inversion is allowed because SOC breaks electronic symmetry (specifically, SU(2) spin rotation symmetry),[86, 87] demonstrating how a combination of symmetries enforces the formation of a zero gap. Because SOC is weak in the bands near the Fermi level, this symmetry is almost preserved, and the zero indirect gap is robust.

We summarize the relevant structural features of $Nb_{11}O_6I_{24}$ in Figure 10a, and compare them to those of monolayer $Nb_3Cl_8$ in Figure 10b. Both materials contain 2D layers of hexagonally packed Nb clusters, creating topological flatbands.[88] The $Nb_3$ clusters in $Nb_3Cl_8$ are identical and achiral, while the $[(Nb_4O)_2(Nb_2O_2)]$ clusters in $Nb_{11}O_6I_{24}$ are chiral. However, monolayer $Nb_3Cl_8$ is noncentrosymmetric (space group *P*3*m*1), whereas $Nb_{11}O_6I_{24}$ is centrosymmetric due to the antiferrochiral arrangement of clusters. The breaking of global inversion symmetry in $Nb_3Cl_8$ opens a gap at the Dirac cones,[88] whereas local symmetry breaking in $Nb_{11}O_6I_{24}$ shifts the VBM and CBM in momentum space while the global symmetries protect their degeneracy.

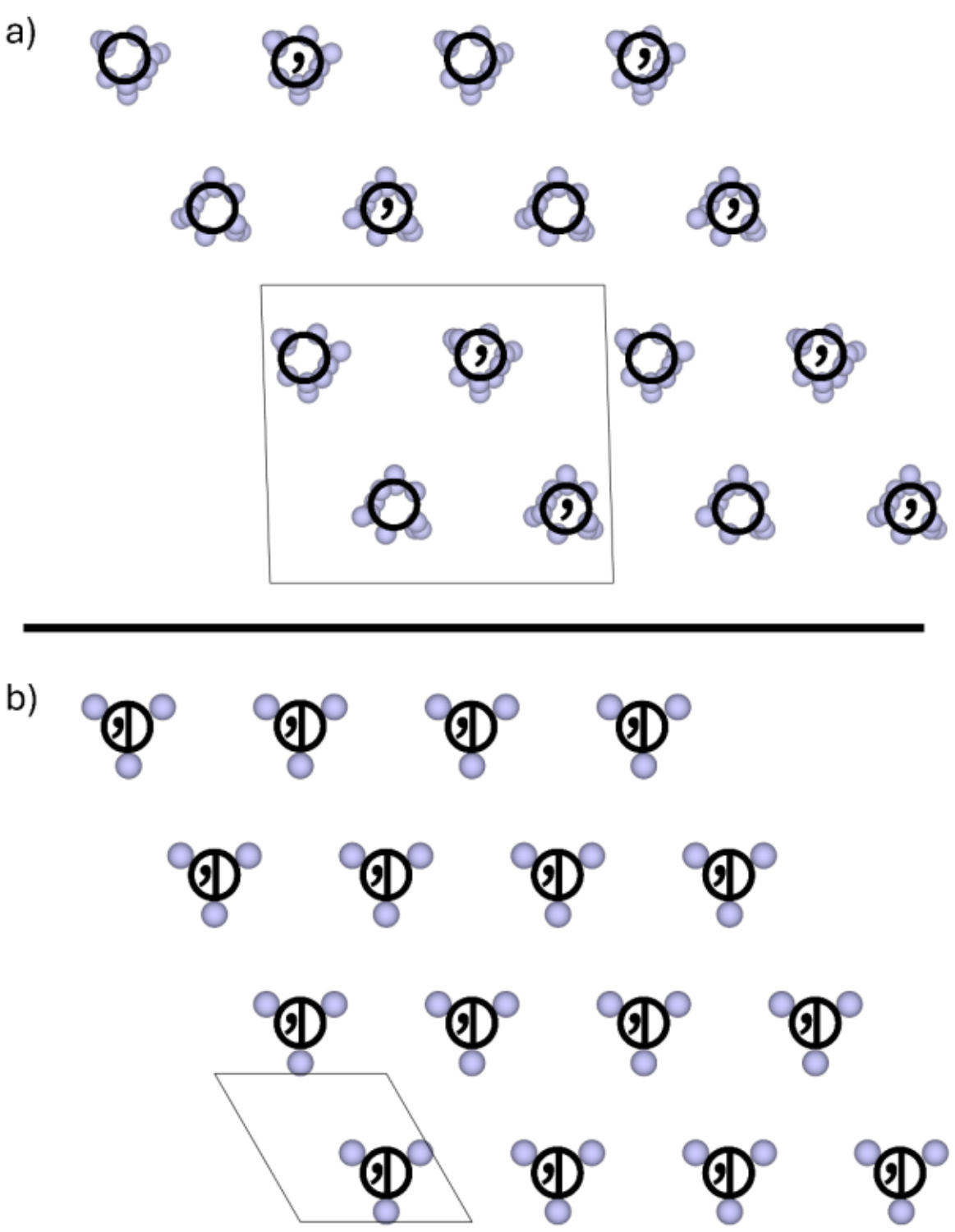

**Figure 10.** Cluster packing and enantiomorphism of $Nb_{11}O_6I_{24}$ (a) and monolayer $Nb_3Cl_8$ (b). Nb atoms are shown as lilac spheres, with the enantiomorphism of the clusters indicated using open circles/circles with commas. Both materials contain a hexagonal packing of clusters; a geometry associated with topological flatbands.[88] The clusters in $Nb_3Cl_8$ are achiral, whereas those in $Nb_{11}O_6I_{24}$ are chiral, providing a local symmetry breaking which removes the momentum-space coincidence of the VBM and CBM. Global inversion symmetry is broken in the breathing Kagome lattice of $Nb_3Cl_8$, creating a direct band gap,[88] and unbroken in $Nb_{11}O_6I_{24}$ by antiferrochirality, preserving its zero gap.

## 2.3 Electrical Conductivity of $Nb_{11}O_6I_{24}$

The two-point probe electrical conductivity, *σ*, of $Nb_{11}O_6I_{24}$ single crystals at 300 K are shown in Figure S10. Exemplifying measurements of two samples are displayed, from which $\sigma$ is calculated as 0.22 S/m for crystal 1 (light blue) and 0.27 S/m for crystal 2 (dark blue). These values are similar to those obtained for $Nb_4OI_{10}$ crystals (0.8–1.2 S/m).[54]

In Figure 11, temperature-dependent two-point probe transport measurements of crystal 1 and 2 in the range of 60 K–300 K are displayed and an Arrhenius-like behavior is observed.[53, 89]

Via linear fitting, activation energies of 0.06 eV for crystal 1 (light blue) and 0.07 eV for crystal 2 (dark blue) were obtained. These results are close to those of $Nb_6I_9S$, having activation energies of 0.048–0.07 eV.[26] We note that the crystals did not exhibit a notable photocurrent under 779 nm excitation, consistent with a small or vanishing band gap.

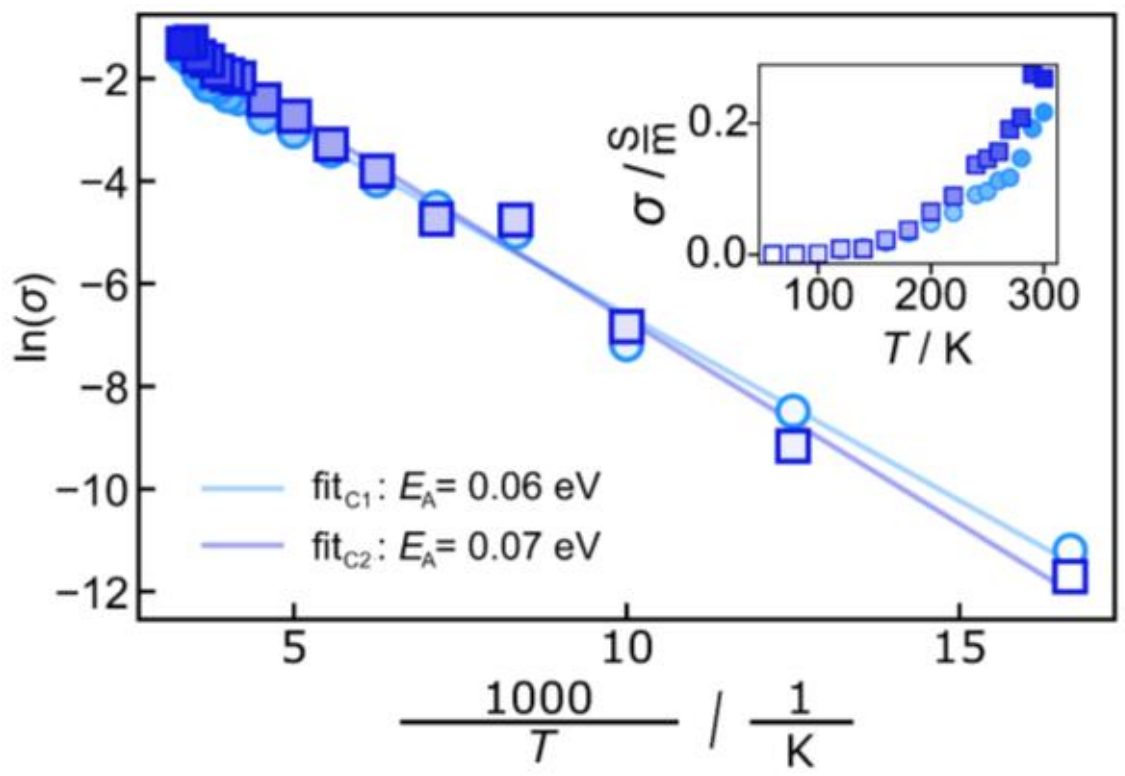

**Figure 11.** Arrhenius plot of the electrical conductivity ($\sigma$ in S/m) from two $Nb_{11}O_6I_{24}$ crystals between 60 K–300 K (in 20 K steps between 60 K–240 K and 240 K–300 K in 10 K steps). The two lines represent linear fits, revealing activation energies of around 0.07 eV (light blue = crystal 1, dark blue = crystal 2). The inset shows the linear plot of conductivity against temperature.

The measured electrical conductivity (Figure 11) cannot by itself confirm the predicted indirect band gap and flat bands, however it is consistent with the DFT calculations. The absence of metallic conduction follows from the DFT-predicted formation of a quantum singlet state, although other consequences of strongly correlated electronic states would require different experimental probes. Direct zero-gap materials often display a non-Arrhenius relationship between conductivity and temperature, with a positive $d\sigma/dT$ gradient at low temperature and a negative gradient at higher temperature.[90, 91] However, the indirect gap predicted in $Nb_{11}O_6I_{24}$ would require momentum transfer from the phononic system for electrical conduction, which would be enhanced upon increasing temperature. Additionally, the flat band structure around the Fermi energy results in a large range of potential transitions with direct and indirect gaps $\leq$ 0.1 eV. The DFT-predicted electronic properties therefore require further investigation and verification. For example, angle-resolved photoemission spectroscopy (ARPES) or magnetoresistance measurements could confirm the unusual features of the band structure. Electronic structure calculations using hybrid exchange–correlation functionals or post-DFT methods, could also be used to corroborate or improve the predicted band gap, however these are highly computationally intensive in such a large system.

## 3. Conclusion

Recently, a number of cluster compounds have been discovered in the field of niobium oxyiodides. Two new niobium oxyiodides are now presented, based on the [$Nb_4O$] butterfly cluster with [NbO] elongations. A butterfly cluster with two elongations is represented by the [$Nb_4O(NbO)_2$] fragment in the structure of $Nb_6O_3I_{15}$, in which the connectivity of cluster generates a three-dimensional network. Two butterfly clusters are connected by a pair of (NbO) units and a bridging (ONbO) unit, leading to the infinite chain structure $^1_\infty$[($Nb_4O$)($ONb_2O$)($Nb_4O$)(ONbO)] in $Nb_{11}O_6I_{24}$.

Electronic structure calculations suggest that $Nb_{11}O_6I_{24}$ has a zero indirect band gap, a highly unusual property not previously reported in an atomic crystal. The zero gap closes a small (0.02 eV) gap between a pair of three-dimensional flatbands which otherwise surround the Fermi energy. The flatbands are shown by examination of the Wannier functions to arise from electron delocalization over [$(Nb_4O)_2(ONb_2O)$] clusters and into the interstitial spaces, where interactions between adjacent clusters lead to quantum interference effects and the formation of a singlet state. Comparison to model systems allows us to suggest that the chirality of the twisted clusters in the overall antiferrochiral crystal leads to the formation of the zero indirect gap.

Electrical conductivity measurements suggest that $Nb_{11}O_6I_{24}$ behaves as a semiconductor with a very small gap, consistent with the flatband picture from DFT.

The sensitive thermal behavior of compounds formed in this heterogeneous reaction of $NbI_4$, $Li_2O$, and $Li_2(CN_2)$ implies metastability and kinetic control of the product formation. While in classical chemical thermodynamics the product formation occurs in or near thermodynamic equilibrium, non-equilibrium systems are characterized by time-dependence, spatial gradients, and fluxes of matter and energy. These features can explain the formation of several metastable compounds[57] and arises when the product formation strongly depends on reaction conditions, particularly on the temperature, the duration at that temperature, and the applied heating and cooling rates. The formation of the compound $Nb_{11}O_6I_{24}$ under such conditions, with its complex crystallographic structure and unusual electronic properties, emphasizes the importance of considering products from such (near) non-equilibrium reaction conditions.

## 4. Experimental Section

All manipulations of starting materials and products were performed in a glovebox under dry argon with moisture and oxygen levels below 1 ppm. $Li_2O$ (ABCR, 95 %) was used as purchased. $NbI_4$ was synthesized as described in the literature.[92] $Li_2(CN_2)$ was synthesized as described previously.[93]

### 4.1 Synthesis

$Nb_6O_3I_{15}$ and $Nb_{11}O_6I_{24}$ were synthesized from $NbI_4$, $Li_2O$ and $Li_2(CN_2)$. For this purpose, $NbI_4$ (160.8 mg, 0.268 mmol), $Li_2O$ (2 mg, 0.067 mmol) and $Li_2(CN_2)$ (7.2 mg, 0.135 mmol) were encapsulated into a fused silica ampoule with 3 cm length and a volume of about 1.5 cm$^3$. The ampule was heated in a box furnace from room temperature to 500 °C with a rate of 0.1 °C/min. The holding time was 1 h before the reaction was cooled down to 450 °C with a rate of 1 °C/min and afterwards to room temperature with a rate of 0.1 °C/min. Block-like crystals of $Nb_6O_3I_{15}$ and elongated, plate-like crystals of $Nb_{11}O_6I_{24}$ were found on the wall of the ampule below the side phases $NbOI_2$ and $NbI_5$ and beside the side phases $Nb_8O_5I_{17}(NbI_5)$ and b-$Nb_4OI_{11}$. At the bottom of the ampule, LiI, $Li_3Nb_7O_5I_{15}$ and amorphous powder can be found. The products are black and quickly decompose in air due to moisture.

### 4.2 Crystallography

A block-like $Nb_6O_3I_{15}$ and a plate-like $Nb_{11}O_6I_{24}$ single-crystal were mounted on a Rigaku XtaLab Synergy-S X-ray diffractometer using Mo-$K_\alpha$ ($\lambda$ = 0.71073 Å) radiation for $Nb_6O_3I_{15}$ and Cu-$K_\alpha$ ($\lambda$ = 1.54184 Å) radiation for $Nb_{11}O_6I_{24}$. The single crystals were kept under $N_2$ cooling at 150 K during the data collection. Corrections for absorption effects were applied with CrysAlisPro 1.171.43.121a (Rigaku Oxford Diffraction, 2022). The crystal structures were solved by the integrated space group and crystal-structure determination routine of SHELXT[94] and full-matrix least-squares refinement with SHELXL-2019/3[94] implemented in Olex2 1.5.[95]

### 4.3 Electrical Conductivity

Conductivity measurements were performed on a Lake Shore Cryotronics CRX-6.5 K probe station with a Keithley 2636B source meter unit. Plate-like crystals of $Nb_{11}O_6I_{24}$ were contacted with silver paste on a silicon substrate with a 770 nm oxide layer and transferred into the measurement chamber under an argon atmosphere. The conductive silver pads at each end of the crystals were connected to the circuit with gold-coated tungsten tips. The chamber was kept under vacuum ($<5\cdot10^{-5}$ mbar) while the temperature was decreased in 10 K steps between 300 K and 240 K and in 20 K steps between 240 K and 60 K during the measurements. Two-point conductivity measurements were performed by varying the applied source-drain voltage from -200 mV to 200 mV while detecting the current. The dimensions (length (L), width (W), height (H)) of the used crystals are for crystal 1: L = 67.2 µm; W = 18.4 µm; H = 16.2 µm and for crystal 2: L = 90.4 µm; W = 16.6 µm; H = 14.4 µm.

### 4.4 Density Functional Theory

Density functional theory (DFT) calculations of the electronic structure of $Nb_{11}O_6I_{24}$ were performed using the Abinit (v. 10)[96] and Quantum Espresso (QE, v. 6.4.1)[97] software packages. The calculation without SOC was repeated with the same DFT input parameters in each software to ensure reproducibility; the calculation including SOC was performed using Abinit. The Perdew–Burke–Ernzerhof exchange–correlation functional[98] was used with the vdw-DFT–D3(BJ) dispersion correction of Grimme.[99] Methfessel–Paxton smearing[100] was used to determine band occupation. Pseudopotentials were used as received from Pseudo Dojo.[101] A 2 × 2 × 2 Monkhorst–Pack grid[102] of k-points was used to sample reciprocal space. Planewave calculations were performed using the projector augmented wave (PAW) formalism,[103] using an energy cutoff of 20 Ha outside of the PAW spheres and a 100 Ha cutoff inside them. These computational parameters were chosen following convergence studies. Structural relaxation was performed prior to the calculation of the band structure. MLWFs were constructed using Wannier90[104] from QE wavefunctions. Topological analysis of the resulting tight-binding band structure was performed using WannierTools.[105] COHP analysis was performed using LOBSTER.[106]

The electronic structure of $Nb_6O_3I_{15}$ was calculated in QE using the same DFT parameters as for $Nb_{11}O_6I_{24}$.

A secondary calculation of the electronic structure of $Nb_{11}O_6I_{24}$ using the LDA[107] exchange–correlation functional was performed in Abinit with a 25 Ha energy cutoff outside of the PAW spheres and a 125 Ha cutoff inside. This calculation showed an essentially identical electronic band structure (see Figure S9), demonstrating that the zero indirect band gap is not an artifact of the choice of functional or pseudopotential.

## SUPPORTING INFORMATION

The Supporting Information contains: Photographs of synthesized crystals, electronic band structure of $Nb_6O_3I_{15}$, niobium, oxygen and iodine character of the band structure of $Nb_{11}O_6I_{24}$, band structure of $Nb_{11}O_6I_{24}$ with tight-binding model, band structure of $Nb_{11}O_6I_{24}$ calculated including SOC, band structure of $Nb_{11}O_6I_{24}$ calculated using LDA, and *I-U* curves of $Nb_{11}O_6I_{24}$

## DATA AVAILABILITY

Crystallographic data have been deposited at the CCDC 2401147 ($Nb_{18}O_9I_{45}$) and 2380623 ($Nb_{11}O_6I_{24}$). Data are available within the article. The data that support the findings of this study are available on request from the corresponding author, H.-J. Meyer. Computational data are available from doi:10.5281/zenodo.18012681.

## AUTHOR INFORMATION

### Corresponding Author

***H.-Jürgen Meyer** – *Section for Solid State and Theoretical Inorganic Chemistry, Institute of Inorganic Chemistry, Eberhard Karls Universität Tübingen, Auf der Morgenstelle 18, 72076 Tübingen, Germany;* E-Mail: *juergen.meyer@uni-tuebingen.de*

***Carl P. Romao** – *Faculty of Nuclear Sciences and Physical Engineering, Czech Technical University in Prague, Břehová 7, 115 19 Prague 1, Czech Republic;* E-Mail: carl.romao@cvut.cz

### Author

**Jan Beitlberger** – *Section for Solid State and Theoretical Inorganic Chemistry, Institute of Inorganic Chemistry, Eberhard Karls Universität Tübingen, Auf der Morgenstelle 18, 72076 Tübingen, Germany*

**Mario Martin** – *Institute for Physical and Theoretical Chemistry, Eberhard Karls Universität Tübingen, Auf der Morgenstelle 18, 72076 Tübingen, Germany*

**Marcus Scheele** – *Institute for Physical and Theoretical Chemistry, Eberhard Karls Universität Tübingen, Auf der Morgenstelle 18, 72076 Tübingen, Germany*

**Marek Matas** – *Faculty of Nuclear Sciences and Physical Engineering, Czech Technical University in Prague, Břehová 7, 115 19 Prague 1, Czech Republic*

**Markus Ströbele** – *Section for Solid State and Theoretical Inorganic Chemistry, Institute of Inorganic Chemistry, Eberhard Karls Universität Tübingen, Auf der Morgenstelle 18, 72076 Tübingen, Germany*

## ACKNOWLEDGMENT

This research was supported by the Deutsche Forschungsgemeinschaft (ME 914-32/1) and the state of Baden-Württemberg through bwHPC and the German Research Foundation (DFG) through grant no INST 40/467-1 FUGG (JUSTUS cluster). C.P.R. acknowledges support from the project FerrMion of the Ministry of Education, Youth and Sports, Czech Republic, co-funded by the European Union (CZ.02.01.01/00/22_008/0004591). M.M. acknowledges support of the Ministry of Education, Youth and Sports of the Czech Republic through e-INFRA CZ (ID:90254) and the support from

the CTU Mobility Project MSCA-F-CZ-III (reg.no. CZ.02.01.01/00/22_010/0008601).

## REFERENCES

(1) Huang, J.-H.; Liu, Y.-J.; Si, Y.; Cui, Y.; Dong, X.-Y.; Zang, S.-Q. Carborane-Cluster-Wrapped Copper Cluster with Cyclodextrin-like Cavities for Chiral Recognition. *J. Am. Chem. Soc.* **2024**, *146* (24), 16729-16736. DOI: 10.1021/jacs.4c04294.

(2) Peña, O. Chevrel phases: Past, present and future. *Physica C, Supercond.* **2015**, *514*, 95-112. DOI: https://doi.org/10.1016/j.physc.2015.02.019.

(3) Gunasekaran, S.; Reed, D. A.; Paley, D. W.; Bartholomew, A. K.; Venkataraman, L.; Steigerwald, M. L.; Roy, X.; Nuckolls, C. Single-Electron Currents in Designer Single-Cluster Devices. *J. Am. Chem. Soc.* **2020**, *142* (35), 14924-14932. DOI: 10.1021/jacs.0c04970.

(4) Hu, J.; Zhang, X.; Hu, C.; Sun, J.; Wang, X.; Lin, H.-Q.; Li, G. Correlated flat bands and quantum spin liquid state in a cluster Mott insulator. *Commun. Phys.* **2023**, *6* (1), 172. DOI: 10.1038/s42005-023-01292-z.

(5) Hernández, J. S.; Guevara, D.; Shamshurin, M.; Benassi, E.; Sokolov, M. N.; Feliz, M. Octahedral Tantalum Bromide Clusters as Catalysts for Light-Driven Hydrogen Evolution. *Inorg. Chem.* **2023**, *62* (46), 19060-19069. DOI: 10.1021/acs.inorgchem.3c03045.

(6) Alonso, J. A. Electronic and Atomic Structure, and Magnetism of Transition-Metal Clusters. *Chem. Rev.* **2000**, *100* (2), 637-678. DOI: 10.1021/cr980391o.

(7) Pauling, L. *The Nature of the Chemical Bond and the Structure of Molecules and Crystals: An Introduction to Modern Structural Chemistry*; Cornell University Press, **1960**.

(8) Simon, A.; von Schnering, H.-G.; Wöhrle, H.; Schäfer, H. Beiträge zur Chemie der Elemente Niob und Tantal. 44. $Nb_6Cl_{14}$ Synthese, Eigenschaften, Struktur. *Z. Anorg. Allg. Chem.* **1965**, *339* (3-4), 155-170. DOI: https://doi.org/10.1002/zaac.19653390306 (acccessed 2024/05/07).

(9) Simon, A.; von Schnering, H.-G.; Schäfer, H. Beiträge zur Chemie der Elemente Niob und Tantal. LXIV. $Nb_6J_{11}$ - Eine Verbindung mit $[Nb_6J_8]$-Gruppen. *Z. Anorg. Allg. Chem.* **1967**, *355* (5-6), 295-310. DOI: https://doi.org/10.1002/zaac.19673550510 (acccessed 2024/05/07).

(10) Ihmaïne, S.; Perrin, C.; Peña, O.; Sergent, M. Structure and magnetic properties of two niobium chlorides with $\lvert Nb_6Cl_{12}\rvert^{n+}$ (*n* = 2, 3) units: $KLuNb_6Cl_{18}$ and $LuNb_6Cl_{18}$. *J. Less-Common Met.* **1988**, *137* (1), 323-332. DOI: https://doi.org/10.1016/0022-5088(88)90097-5.

(11) von Schnering, H.-G.; Wöhrle, H.; Schäfer, H. Die Kristallstruktur der Verbindung $Nb_3Cl_8$. *Sci. Nat.* **1961**, *48* (6), 159-159. DOI: 10.1007/BF00639538.

(12) Simon, A.; von Schnering, H.-G. *β*-$Nb_3Br_8$ und *β*-$Nb_3J_8$ Darstellung, Eigenschaften und Struktur. *J. Less-Common Met.* **1966**, *11* (1), 31-46. DOI: https://doi.org/10.1016/0022-5088(66)90054-3.

(13) Ehrlich, P.; Plöger, F.; Pietzka, G. Über Niobtrifluorid. *Z. Anorg. Allg. Chem.* **1955**, *282* (1-6), 19-23. DOI: https://doi.org/10.1002/zaac.19552820105 (acccessed 2025/10/13).

(14) Seabaugh, P. W.; Corbett, J. D. The Niobium Iodides. Characterization of Niobium (IV) Iodide, Niobium (III) Iodide, and Triniobium Octaiodide. *Inorg. Chem.* **1965**, *4* (2), 176-181.

(15) Schäfer, H.; von Schnering, H.-G.; Niehues, K. J.; Nieder-Vahrenholz, H. G. Beiträge zur Chemie der Elemente Niob und Tantal: XLVII. Niobfluoride. *J. Less-Common Met.* **1965**, *9* (2), 95-104. DOI: https://doi.org/10.1016/0022-5088(65)90087-1.

(16) Taylor, D. R.; Calabrese, J. C.; Larsen, E. M. Crystal Structure of Niobium Tetrachloride. *Inorg. Chem.* **1977**, *16* (3), 721-722. DOI: 10.1021/ic50169a048.

(17) Benjamin, S. L.; Chang, Y.-P.; Hector, A. L.; Jura, M.; Levason, W.; Reid, G.; Stenning, G. Niobium tetrahalide complexes with neutral diphosphine ligands. *Dalton Trans.* **2016**, *45* (19), 8192-8200.

(18) Dahl, L. F.; Wampler, D. L. The Crystal Structure of *α*-Niobium Tetraiodide. *Acta Crystallog.* **1962**, *15* (9), 903-911. DOI: doi:10.1107/S0365110X62002340.

(19) Edwards, A. J. 717. The Structures of Niobium and Tantalum Pentafluorides. *J. Chem. Soc.* **1964**, 3714-3718, 10.1039/JR9640003714. DOI: 10.1039/jr9640003714.

(20) Hönle, W.; von Schnering, H.-G. Crystal structure of niobium pentachloride, $NbCl_5$. *Z. Kristallogr.* **1990**, *191* (1-2), 139-140. DOI: 10.1524/zkri.1990.191.1-2.139 (acccessed 2025-03-18).

(21) Müller, U.; Klingelhöfer, P. *β*-$NbBr_5$, eine neue Modifikation von Niobpentabromid mit einer eindimensionalen Lagenfehlordnung. **1983**, *38* (5), 559-561. DOI: doi:10.1515/znb-1983-0506 (acccessed 2025-10-13).

(22) Littke, W.; Brauer, G. Darstellung und Kristallstruktur von Niobpentajodid. *Z. Anorg. Allg. Chem.* **1963**, *325* (3-4), 122-129. DOI: https://doi.org/10.1002/zaac.19633250304 (acccessed 2025/10/13).

(23) Krebs, B.; Sinram, D. Darstellung, Struktur und Eigenschaften einer neuen Modifikation von $NbI_5$. *Z. Naturforsch. B* **1980**, *35* (1), 12-16.

(24) Grytsiuk, S.; Katsnelson, M. I.; Loon, E. G. C. P. v.; Rösner, M. $Nb_3Cl_8$: a prototypical layered Mott-Hubbard insulator. *npj Quantum Mater.* **2024**, *9* (1), 8. DOI: 10.1038/s41535-024-00619-5.

(25) Carta, A.; Mlkvik, P.; Grahlow, F.; Ströbele, M.; Meyer, H.-J.; Romao, C. P.; Spaldin, N. A.; Ederer, C. Hubbard dimer physics and the magnetostructural transition in the correlated cluster material $Nb_3Cl_8$. *arXiv preprint arXiv:2509.03988* **2025**.

(26) Meyer, H.-J.; Corbett, J. D. Synthesis and Structure of the Novel Chain Compound Niobium Iodide Sulfide $Nb_6I_9S$ and Its Hydride. *Inorg. Chem.* **1991**, *30* (5), 963-967. DOI: 10.1021/ic00005a017.

(27) Pasco, C. M. Electronic and magnetic properties of layered two-dimensional materials. Dissertation, Johns Hopkins University, 2021. https://jscholarship.library.jhu.edu/server/api/core/bitstreams/10ba6e2f-c7c6-4c1b-b1fa-0684bed80962/content.

(28) Khvorykh, G. V.; Shevelkov, A. V.; Dolgikh, V. A.; Popovkin, B. A. Niobium Thiobromide, $Nb_3SBr_7$, with Triangle $Nb_3$ Cluster: Structure and Bonding. *J. Solid State Chem.* **1995**, *120* (2), 311-315. DOI: 10.1006/jssc.1995.1413.

(29) Schmidt, P. J.; Thiele, G. A New Structural Variation of $Nb_3YX_7$ Compounds: Monoclinic $Nb_3SI_7$. *Acta Crystallogr. Sect. C* **1997**, *53* (12), 1743-1745. DOI: doi:10.1107/S0108270197008196.

(30) Meyer, H.-J. $CsNb_3Br_7S$: Synthese, Struktur und Bindungsverhältnisse. *Z. Anorg. Allg. Chem.* **1994**, *620* (5), 863-866. DOI: https://doi.org/10.1002/zaac.19946200519 (acccessed 2025/01/08).

(31) Grahlow, F.; Strauß, F.; Scheele, M.; Ströbele, M.; Carta, A.; Weber, S. F.; Kroeker, S.; Romao, C. P.; Meyer, H.-J. Electronic structure and transport in the potential Luttinger liquids $CsNb_3Br_7S$ and $RbNb_3Br_7S$. *Phys. Chem. Chem. Phys.* **2024**, *26* (15), 11789-11797.

(32) Seela, J. L.; Huffman, J. C.; Christou, G. The First Example of a Niobium–Sulphide–Thiolate Cluster: Metal–Metal Bonding and $\mu_4$-Sulphide Groups in Tetranuclear $[Nb_4S_2(SPh)_{12}]^{4-}$. *J. Chem. Soc., Chem. Commun.* **1987**, (16), 1258-1260.

(33) Cotton, F. A.; Shang, M. New Niobium Complexes with Alkynes. 2. Tetranuclear Compounds with Niobium-Niobium Bonds, an Unprecedented Type of Tetracarbon Ligand, and Oxygen in a Rectangular Environment. *J. Am. Chem. Soc.* **1990**, *112* (4), 1584-1590.

(34) Broll, A.; Simon, A.; von Schnering, H.-G.; Schäfer, H. Beiträge zur Chemie der Elemente Niob und Tantal. LXXIII. $CsNb_4Cl_{11}$, $CsNb_4Br_{11}$ und $RbNb_4Cl_{11}$ Verbindungen mit planaren $Nb_4$-Gruppen. *Z. Anorg. Allg. Chem.* **1969**, *367* (1-2), 1-18.

(35) Yaich, H. B.; Jegaden, J. C.; Potel, M.; Sergent, M.; Rastogi, A. K.; Tournier, R. Nouveaux chalcogénures et chalcohalogénures à clusters tétraédriques $Nb_4$ ou $Ta_4$. *J. Less-Common Met.* **1984**, *102* (1), 9-22.

(36) Ströbele, M.; Oeckler, O.; Thelen, M.; Fink, R. F.; Krishnamurthy, A.; Kroeker, S.; Meyer, H.-J. Pnictide-Capped Butterfly Cluster in the Crystal Structure of $Nb_4PnX_{11}$ (*Pn*= N, P; *X*= Cl, Br, I). *Inorg. Chem.* **2022**, *61* (44), 17599-17608.

(37) Peng, Y.; Powell, A. K. What do 3*d*-4*f* butterflies tell us? *Coord. Chem. Rev.* **2021**, *426*, 213490. DOI: https://doi.org/10.1016/j.ccr.2020.213490.

(38) Ziebarth, R. P.; Corbett, J. D. Centered Zirconium Chloride Cluster Compounds with the $Ta_6Cl_{15}$ Structure. *J. Less-Common Met.* **1988**, *137* (1), 21-34. DOI: https://doi.org/10.1016/0022-5088(88)90072-0.

(39) Ziebarth, R. P.; Corbett, J. D. New Zirconium Chloride Cluster Phases with the Stoichiometries $Zr_6Cl_{12}Z$ and $Zr_6Cl_{14}Z$ That Are Stabilized by Interstitial Atoms ($Z$ = H, Be, B, C). *J. Solid State Chem.* **1989**, *80* (1), 56-67. DOI: https://doi.org/10.1016/0022-4596(89)90031-5.

(40) Zhang, J.; Corbett, J. D. $Cs_3Zr_7Cl_{20}Mn$: A Zirconium Cluster Network Compound with Isolated $ZrCl_5^-$ Units in a Stuffed Perovskite Structure. *Inorg. Chem.* **1995**, *34* (7), 1652-1656. DOI: 10.1021/ic00111a008.

(41) Zhang, J.; Corbett, J. D. New Cluster Phases $A^{I}Zr_6Cl_{14}Mn$ and $Zr_6Cl_{14}Fe$: A Second Structure Type with Small Cations. *J. Solid State Chem.* **1994**, *109* (2), 265-271. DOI: https://doi.org/10.1006/jssc.1994.1102.

(42) Zhang, J.; Corbett, J. D. Two Families of Centered Zirconium Cluster Phases with $M_{1,2}M'Cl_6\cdot Zr_6Cl_{12}Z$ Compositions. *Inorg. Chem.* **1993**, *32* (9), 1566-1572. DOI: 10.1021/ic00061a009.

(43) Manassero, M.; Sansoni, M.; Longoni, G. Crystal Structure of $[Me_3NCH_2Ph][Fe_4(CO)_{13}H]$. A 'Butterfly' Metal Cluster with an Unusually Bonded Carbonyl Group. *J. Chem. Soc., Chem. Commun.* **1976**, (22), 919-920.

(44) Bradley, J. S.; Ansell, G. B.; Hill, E. W. Homogeneous Carbon Monoxide Hydrogenation on Multiple Sites: A Dissociative Pathway to Oxygenates. *J. Am. Chem. Soc.* **1979**, *101* (24), 7417-7419.

(45) Bradley, J. S.; Ansell, G. B.; Leonowicz, M. E.; Hill, E. W. Synthesis and Molecular Structure of $\mu^4$-Carbido-$\mu^2$-carbonyl-dodecacarbonyltetrairon, a Neutral Iron Butterfly Cluster Bearing an Exposed Carbon Atom. *J. Am. Chem. Soc.* **1981**, *103* (16), 4968-4970.

(46) Chisholm, M. H.; Folting, K.; Huffman, J. C.; Leonelli, J.; Marchant, N. S.; Smith, C. A.; Taylor, L. C. E. Tetranuclear Carbidotungsten and Nitridomolybdenum Clusters Supported by Alkoxide Ligands: $W_4(C)(O\text{-}i\text{-}Pr)_{12}(NMe)$ and $Mo_4(N)_2(O\text{-}i\text{-}Pr)_{12}$. *J. Am. Chem. Soc.* **1985**, *107* (12), 3722-3724.

(47) Chisholm, M. H.; Errington, R. J.; Folting, K.; Huffman, J. C. Square and Butterfly, 12-Electron Molybdenum ($Mo_4$) Clusters Formed by Coupling Molybdenum-Molybdenum Triple Bonds. *J. Am. Chem. Soc.* **1982**, *104* (7), 2025-2027.

(48) Grahlow, F.; Strauß, F.; Schmidt, P.; Valenta, J.; Ströbele, M.; Scheele, M.; Romao, C. P.; Meyer, H.-J. $Ta_4SBr_{11}$: A Cluster Mott Insulator with a Corrugated, Van der Waals Layered Structure. *Inorg. Chem.* **2024**, *63* (42), 19717-19727. DOI: 10.1021/acs.inorgchem.4c02896.

(49) Ströbele, M.; Meyer, H.-J. Synthesen, Kristallstrukturen und magnetische Eigenschaften von $[Li(12\text{-}Krone\text{-}4)_2][Li(12\text{-}Krone\text{-}4)(OH_2)]_2[Nb_6Cl_{18}]$, $[Li(15\text{-}Krone\text{-}5)_2(OH_2)]_3[Nb_6Cl_{18}]$ und $[(18\text{-}Krone\text{-}6)_2(O_2H_5)]_3[Nb_6Cl_{18}]$. *Z. Naturforsch. B* **2001**, *56* (10), 1025-1034. DOI: doi:10.1515/znb-2001-1011 (acccessed 2025-10-20).

(50) Cotton, F. A.; Diebold, M. P.; Feng, X.; Roth, W. J. Discrete Trinuclear Complexes of Niobium and Tantalum Related to the Local Structure in $Nb_3Cl_8$. *Inorg. Chem.* **1988**, *27* (19), 3413-3421. DOI: 10.1021/ic00292a029.

(51) Duraisamy, T.; Lachgar, A. $A$VNb$_3$Cl$_{11}$ ($A$ = K, Rb, Cs, Tl): A Series of Layered Vanadium Niobium Halides Based on Triangular $Nb_3$ Clusters. *Inorg. Chem.* **2003**, *42* (24), 7747-7751. DOI: 10.1021/ic0341439.

(52) Kennedy, J. R.; Simon, A. Chemical Intercalation of Sodium into $\alpha$-$Nb_3Cl_8$. *Inorg. Chem.* **1991**, *30* (11), 2564-2567. DOI: 10.1021/ic00011a020.

(53) Beitlberger, J.; Martin, M.; Scheele, M.; Schmidt, P.; Ströbele, M.; Meyer, H.-J. The family of tetranuclear $Nb_4OI_{12-x}$ clusters ($x$ = 0, 1, 2): from the molecular $Nb_4OI_{12}$ cluster to extended chains and layers. *Dalton Trans.* **2025**, *54* (13), 5486-5493. DOI: 10.1039/D5DT00174A.

(54) Beitlberger, J.; Ströbele, M.; Strauß, F.; Scheele, M.; Romao, C. P.; Meyer, H.-J. The Rectangular Niobium Oxyiodide Cluster $Nb_4OI_{10}$ - A Narrow Band-Gap Semiconductor. *Eur. J. Inorg. Chem.* **2024**, *27* (28). DOI: 10.1002/ejic.202400329.

(55) Grahlow, F.; Beitlberger, J.; Martin, M.; Juriatti, E.; Peisert, H.; Scheele, M.; Ströbele, M.; Romao, C. P.; Meyer, H.-J. Structural modifications of $M_5O_4I_{11}$ ($M$ = Nb, Ta) cluster networks from heterogeneous solid-state reactions. *Dalton Trans.* **2025**, *54* (44), 16593-16604, 10.1039/D5DT02097B. DOI: 10.1039/d5dt02097b From NLM PubMed-not-MEDLINE.

(56) Beitlberger, J.; Ströbele, M.; Schmidt, P.; Romao, C. P.; Meyer, H.-J. Niobium oxyiodide cluster compounds $Li_3Nb_7O_5I_{15}$ and $Nb_8O_5I_{17}(NbI_5)$ with expanding cluster architectures and multicentre Nb–Nb bonding. *Dalton Trans.* **2025**, *54* (38), 14376-14383. DOI: 10.1039/D5DT01819F.

(57) Lebon, G.; Jou, D.; Casas-Vázquez, J. *Understanding non-equilibrium thermodynamics: foundations, applications, frontiers*; Springer, **2008**.

(58) Haraguchi, Y.; Michioka, C.; Ishikawa, M.; Nakano, Y.; Yamochi, H.; Ueda, H.; Yoshimura, K. Magnetic–Nonmagnetic Phase Transition with Interlayer Charge Disproportionation of $Nb_3$ Trimers in the Cluster Compound $Nb_3Cl_8$. *Inorg. Chem.* **2017**, *56* (6), 3483-3488. DOI: 10.1021/acs.inorgchem.6b03028.

(59) Tremel, W. $Nb_4OTe_9I_4$: a One-dimensional Chain Compound containing Tetranuclear Oxygen-centred Niobium Clusters. *J. Chem. Soc., Chem. Commun.* **1992**, (9), 709-710, 10.1039/C39920000709. DOI: 10.1039/C39920000709.

(60) Wang, X.-L.; Dou, S. X.; Zhang, C. Zero-gap materials for future spintronics, electronics and optics. *NPG Asia Mater.* **2010**, *2* (1), 31-38. DOI: 10.1038/asiamat.2010.7.

(61) Palumbo, G. Topological Phase Transitions with Zero Indirect Band Gaps. *J. Phys.: Condens. Matter* **2024**, *36* (26), 26LT01. DOI: 10.1088/1361-648X/ad3872.

(62) Palumbo, G. Zero Indirect Band Gap in Non-Hermitian Systems. *arXiv preprint arXiv:2509.15102* **2025**.

(63) Andrijauskas, T.; Anisimovas, E.; Račiūnas, M.; Mekys, A.; Kudriašov, V.; Spielman, I. B.; Juzeliūnas, G. Three-level Haldane-like model on a dice optical lattice. *Phys. Rev. A* **2015**, *92* (3), 033617. DOI: 10.1103/PhysRevA.92.033617.

(64) Palumbo, G.; Meichanetzidis, K. Two-dimensional Chern semimetals on the Lieb lattice. *Phys. Rev. B* **2015**, *92* (23), 235106. DOI: 10.1103/PhysRevB.92.235106.

(65) Krishtopenko, S. S.; Yahniuk, I.; But, D. B.; Gavrilenko, V. I.; Knap, W.; Teppe, F. Pressure- and temperature-driven phase transitions in HgTe quantum wells. *Phys. Rev. B* **2016**, *94* (24), 245402. DOI: 10.1103/PhysRevB.94.245402.

(66) Pyrialakos, G. G.; Apostolidis, A.; Khajavikhan, M.; Kantartzis, N. V.; Christodoulides, D. N. Antichiral topological phases and protected bulk transport in dual-helix Floquet lattices. *Phys. Rev. B* **2023**, *107* (17), 174313. DOI: 10.1103/PhysRevB.107.174313.

(67) Chen, J.; Zheng, Y.; Yang, S.; Alù, A.; Li, Z.-Y.; Qiu, C.-W. Chern-Protected Flatband Edge State in Metaphotonics. *Phys. Rev. Lett.* **2025**, *134* (22), 223806. DOI: 10.1103/PhysRevLett.134.223806.

(68) Hinuma, Y.; Pizzi, G.; Kumagai, Y.; Oba, F.; Tanaka, I. Band structure diagram paths based on crystallography. *Comput. Mater. Sci.* **2017**, *128*, 140-184.

(69) Cao, Y.; Fatemi, V.; Fang, S.; Watanabe, K.; Taniguchi, T.; Kaxiras, E.; Jarillo-Herrero, P. Unconventional superconductivity in magic-angle graphene superlattices. *Nat.* **2018**, *556* (7699), 43-50. DOI: 10.1038/nature26160.

(70) Tang, E.; Mei, J.-W.; Wen, X.-G. High-Temperature Fractional Quantum Hall States. *Phys. Rev. Lett.* **2011**, *106* (23), 236802. DOI: 10.1103/PhysRevLett.106.236802.

(71) Wang, H.-C.; Rauch, T.; Tellez-Mora, A.; Wirtz, L.; Romero, A. H.; Marques, M. A. L. Exploring flat-band properties in two-dimensional $M_3QX_7$ compounds. *Phys. Chem. Chem. Phys.* **2024**, *26* (32), 21558-21567, 10.1039/D4CP01196A. DOI: 10.1039/D4CP01196A.

(72) Wu, S. L.; Ren, Z.-H.; Yang, L.; Wang, M.-Y.; Zhang, X.-P.; Fan, X.-Y.; Zhang, H.-C.; Li, X.; Wang, G.; Wang, C.; et al. Efficiency-Tunable Field-Free Josephson Diode Effect in Nb3Cl8 Based van der Waals Junctions. *Nano Letters* **2025**, *25* (51), 17619-17627. DOI: 10.1021/acs.nanolett.5c04336.

(73) Dubbelman, M. P.; Wu, H.; Aretz, J.; Wang, Y.; Pasco, C. M.; Zhao, Y.; Kyrk, T. M.; Yang, J.; Xu, X.; McQueen, T. M. Driving the field-free Josephson diode effect using Kagome Mott insulator barriers. *arXiv preprint arXiv:2512.17099* **2025**.

(74) Aoki, H. Flat bands in condensed-matter systems – perspective for magnetism and superconductivity. *Contemp. Phys.* **2025**, *66* (1-4), 1-38. DOI: 10.1080/00107514.2025.2550105.

(75) Marzari, N.; Mostofi, A. A.; Yates, J. R.; Souza, I.; Vanderbilt, D. Maximally localized Wannier functions: Theory and applications. *Rev. Mod. Phys.* **2012**, *84* (4), 1419-1475. DOI: 10.1103/RevModPhys.84.1419.

(76) Deringer, V. L.; Tchougréeff, A. L.; Dronskowski, R. Crystal Orbital Hamilton Population (COHP) Analysis As Projected from Plane-Wave Basis Sets. *The Journal of Physical Chemistry A* **2011**, *115* (21), 5461-5466. DOI: 10.1021/jp202489s.

(77) Dronskowski, R.; Bloechl, P. E. Crystal orbital Hamilton populations (COHP): energy-resolved visualization of chemical bonding in solids based on density-functional calculations. *The Journal of Physical Chemistry* **1993**, *97* (33), 8617-8624. DOI: 10.1021/j100135a014.

(78) Kang, M.; Fang, S.; Ye, L.; Po, H. C.; Denlinger, J.; Jozwiak, C.; Bostwick, A.; Rotenberg, E.; Kaxiras, E.; Checkelsky, J. G.; Comin, R. Topological flat bands in frustrated kagome lattice CoSn. *Nat. Commun.* **2020**, *11* (1), 4004. DOI: 10.1038/s41467-020-17465-1.

(79) Sun, K.; Gu, Z.; Katsura, H.; Das Sarma, S. Nearly Flatbands with Nontrivial Topology. *Phys. Rev. Lett.* **2011**, *106* (23), 236803. DOI: 10.1103/PhysRevLett.106.236803.

(80) Kane, C. L.; Mele, E. J. $\mathbb{Z}_2$ Topological Order and the Quantum Spin Hall Effect. *Phys. Rev. Lett.* **2005**, *95* (14), 146802. DOI: 10.1103/PhysRevLett.95.146802.

(81) Yu, R.; Qi, X. L.; Bernevig, A.; Fang, Z.; Dai, X. Equivalent expression of $\mathbb{Z}_2$ topological invariant for band insulators using the non-Abelian Berry connection. *Phys. Rev. B* **2011**, *84* (7), 075119. DOI: 10.1103/PhysRevB.84.075119.

(82) Fu, L.; Kane, C. L.; Mele, E. J. Topological Insulators in Three Dimensions. *Phys. Rev. Lett.* **2007**, *98* (10), 106803. DOI: 10.1103/PhysRevLett.98.106803.

(83) Zhong, J.; Yang, M.; Shi, Z.; Li, Y.; Mu, D.; Liu, Y.; Cheng, N.; Zhao, W.; Hao, W.; Wang, J.; et al. Towards layer-selective quantum spin hall channels in weak topological insulator $Bi_4Br_2I_2$. *Nat. Commun.* **2023**, *14* (1), 4964. DOI: 10.1038/s41467-023-40735-7.

(84) Rademaker, L.; Protopopov, I. V.; Abanin, D. A. Topological flat bands and correlated states in twisted monolayer-bilayer graphene. *Phys. Rev. Res.* **2020**, *2* (3), 033150. DOI: 10.1103/PhysRevResearch.2.033150.

(85) Wang, Y.; Li, Y.-X.; Cseh, L.; Chen, Y.-X.; Yang, S.-G.; Zeng, X.; Liu, F.; Hu, W.; Ungar, G. Enantiomers Self-Sort into Separate Counter-Twisted Ribbons of the *Fddd* Liquid Crystal─Antiferrochirality and Parachirality. *J. Am. Chem. Soc.* **2023**, *145* (31), 17443-17460. DOI: 10.1021/jacs.3c06164.

(86) Island, J. O.; Cui, X.; Lewandowski, C.; Khoo, J. Y.; Spanton, E. M.; Zhou, H.; Rhodes, D.; Hone, J. C.; Taniguchi, T.; Watanabe, K.; et al. Spin–orbit-driven band inversion in bilayer graphene by the van der Waals proximity effect. *Nat.* **2019**, *571* (7763), 85-89. DOI: 10.1038/s41586-019-1304-2.

(87) Kane, C. L.; Mele, E. J. Quantum Spin Hall Effect in Graphene. *Phys. Rev. Lett.* **2005**, *95* (22), 226801. DOI: 10.1103/PhysRevLett.95.226801.

(88) Sun, Z.; Zhou, H.; Wang, C.; Kumar, S.; Geng, D.; Yue, S.; Han, X.; Haraguchi, Y.; Shimada, K.; Cheng, P.; et al. Observation of Topological Flat Bands in the Kagome Semiconductor Nb3Cl8. *Nano Letters* **2022**, *22* (11), 4596-4602. DOI: 10.1021/acs.nanolett.2c00778.

(89) Al-Fa'ouri, A. M.; Lafi, O. A.; Abu-Safe, H. H.; Abu-Kharma, M. Investigation of optical and electrical properties of copper oxide-polyvinyl alcohol nanocomposites for solar cell applications. *Arab. J. Chem.* **2023**, *16* (4), 104535. DOI: ARTN 104535
10.1016/j.arabjc.2022.104535.

(90) Tajima, N.; Kajita, K. Experimental study of organic zero-gap conductor α-$(BEDT\text{-}TTF)_2I_3$. *Sci. Technol. Adv. Mater.* **2009**, *10* (2), 024308. DOI: 10.1088/1468-6996/10/2/024308.

(91) Müller, M.; Bräuninger, M.; Trauzettel, B. Temperature Dependence of the Conductivity of Ballistic Graphene. *Phys. Rev. Lett.* **2009**, *103* (19), 196801. DOI: 10.1103/PhysRevLett.103.196801.

(92) Brauer, G. *Handbuch der präparativen anorganischen Chemie*; Enke, **1975**.

(93) Srinivasan, R.; Ströbele, M.; Meyer, H.-J. Chains of [$RE_6$] Octahedra Coupled by (NCN) Links in the Network Structure of $RE_2Cl(CN_2)N$. Synthesis and Structure of Two Novel Rare Earth Chloride Carbodiimide Nitrides with Structures Related to the $RE_2Cl_3$ Type. *Inorg. Chem.* **2003**, *42* (11), 3406-3411. DOI: 10.1021/ic020685z.

(94) Sheldrick, G. M. Crystal structure refinement with SHELXL. *Acta Crystallogr., Sect. C:Struct. Chem.* **2015**, *71* (1), 3-8.

(95) Dolomanov, O. V.; Bourhis, L. J.; Gildea, R. J.; Howard, J. A. K.; Puschmann, H. OLEX2: a complete structure solution, refinement and analysis program. *J. Appl. Crystallogr.* **2009**, *42* (2), 339-341.

(96) Verstraete, M. J.; Abreu, J.; Allemand, G. E.; Amadon, B.; Antonius, G.; Azizi, M.; Baguet, L.; Barat, C.; Bastogne, L.; Béjaud, R.; et al. Abinit 2025: New capabilities for the predictive modeling of solids and nanomaterials. *J. Chem. Phys.* **2025**, *163* (16), 164126. DOI: 10.1063/5.0288278 (acccessed 2/5/2026).

(97) Giannozzi, P.; Andreussi, O.; Brumme, T.; Bunau, O.; Buongiorno Nardelli, M.; Calandra, M.; Car, R.; Cavazzoni, C.; Ceresoli, D.; Cococcioni, M.; et al. Advanced capabilities for materials modelling with Quantum ESPRESSO. *J. Phys.: Condens. Matter* **2017**, *29* (46), 465901. DOI: 10.1088/1361-648X/aa8f79.

(98) Perdew, J. P.; Burke, K.; Ernzerhof, M. Generalized Gradient Approximation Made Simple. *Phys. Rev. Lett.* **1996**, *77* (18), 3865-3868. DOI: 10.1103/PhysRevLett.77.3865.

(99) Grimme, S.; Antony, J.; Ehrlich, S.; Krieg, H. A consistent and accurate *ab initio* parametrization of density functional dispersion correction (DFT-D) for the 94 elements H-Pu. *J. Chem. Phys.* **2010**, *132* (15), 154104. DOI: 10.1063/1.3382344 (acccessed 4/16/2024).

(100) Methfessel, M.; Paxton, A. T. High-precision sampling for Brillouin-zone integration in metals. *Phys. Rev. B* **1989**, *40* (6), 3616-3621. DOI: 10.1103/PhysRevB.40.3616.

(101) van Setten, M. J.; Giantomassi, M.; Bousquet, E.; Verstraete, M. J.; Hamann, D. R.; Gonze, X.; Rignanese, G. M. The PseudoDojo: Training and grading a 85 element optimized norm-conserving pseudopotential table. *Comput. Phys. Commun.* **2018**, *226*, 39-54. DOI: https://doi.org/10.1016/j.cpc.2018.01.012.

(102) Monkhorst, H. J.; Pack, J. D. Special points for Brillouin-zone integrations. *Phys. Rev. B* **1976**, *13* (12), 5188-5192. DOI: 10.1103/PhysRevB.13.5188.

(103) Torrent, M.; Jollet, F.; Bottin, F.; Zérah, G.; Gonze, X. Implementation of the projector augmented-wave method in the ABINIT code: Application to the study of iron under pressure. *Comput. Mater. Sci.* **2008**, *42* (2), 337-351. DOI: https://doi.org/10.1016/j.commatsci.2007.07.020.

(104) Mostofi, A. A.; Yates, J. R.; Lee, Y.-S.; Souza, I.; Vanderbilt, D.; Marzari, N. wannier90: A tool for obtaining maximally-localised Wannier functions. *Comput. Phys. Commun.* **2008**, *178* (9), 685-699. DOI: https://doi.org/10.1016/j.cpc.2007.11.016.

(105) Wu, Q.; Zhang, S.; Song, H.-F.; Troyer, M.; Soluyanov, A. A. WannierTools: An open-source software package for novel topological materials. *Comput. Phys. Commun.* **2018**, *224*, 405-416. DOI: https://doi.org/10.1016/j.cpc.2017.09.033.

(106) Nelson, R.; Ertural, C.; George, J.; Deringer, V. L.; Hautier, G.; Dronskowski, R. LOBSTER: Local orbital projections, atomic charges, and chemical-bonding analysis from projector-augmented-wave-based density-functional theory. *Journal of computational chemistry* **2020**, *41* (21), 1931-1940.

(107) Perdew, J. P.; Wang, Y. Accurate and simple analytic representation of the electron-gas correlation energy. *Phys. Rev. B* **1992**, *45* (23), 13244-13249. DOI: 10.1103/PhysRevB.45.13244.

# Zero Indirect Band Gap and Flat Bands in a Niobium Oxyiodide Cluster Material – Supporting Information

Jan Beitlberger,[a] Mario Martin,[b] Marcus Scheele,[b] Marek Matas,[c] Carl P. Romao,[*c] Markus Ströbele,[a] and H.-Jürgen Meyer[*a]

[a] Section for Solid State and Theoretical Inorganic Chemistry, Institute of Inorganic Chemistry, Auf der Morgenstelle 18, 72076 Tübingen, Germany.

[b] Institute for Physical and Theoretical Chemistry, Eberhard-Karls-Universität Tübingen, Auf der Morgenstelle 18, 72076 Tübingen, Germany.

[c] Faculty of Nuclear Sciences and Physical Engineering, Czech Technical University in Prague, Czech Republic.

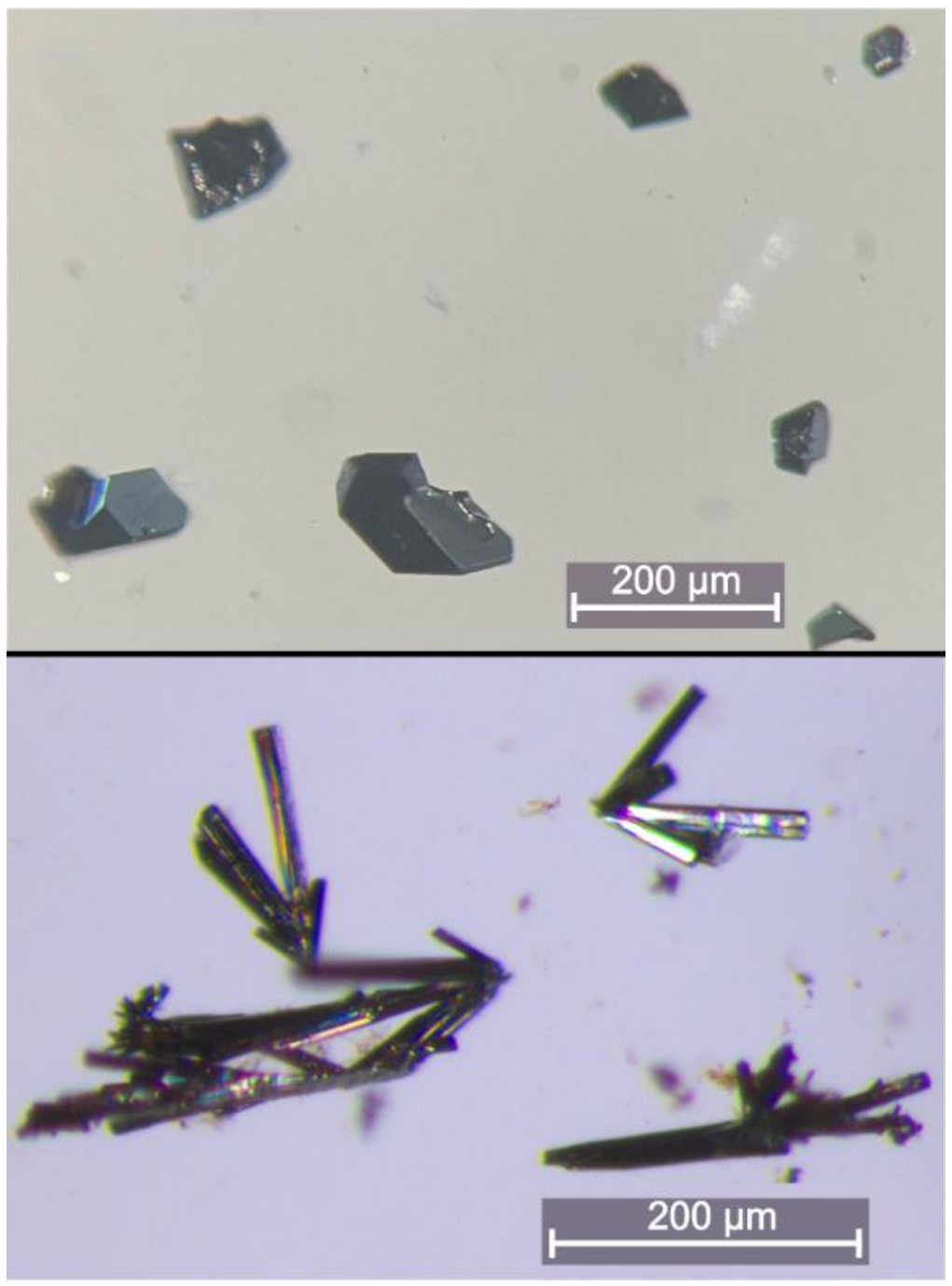

Fig. S1: Examples of block-shaped $Nb_6O_3I_{15}$ (top) and columnar $Nb_{11}O_6I_{24}$ (bottom) crystals

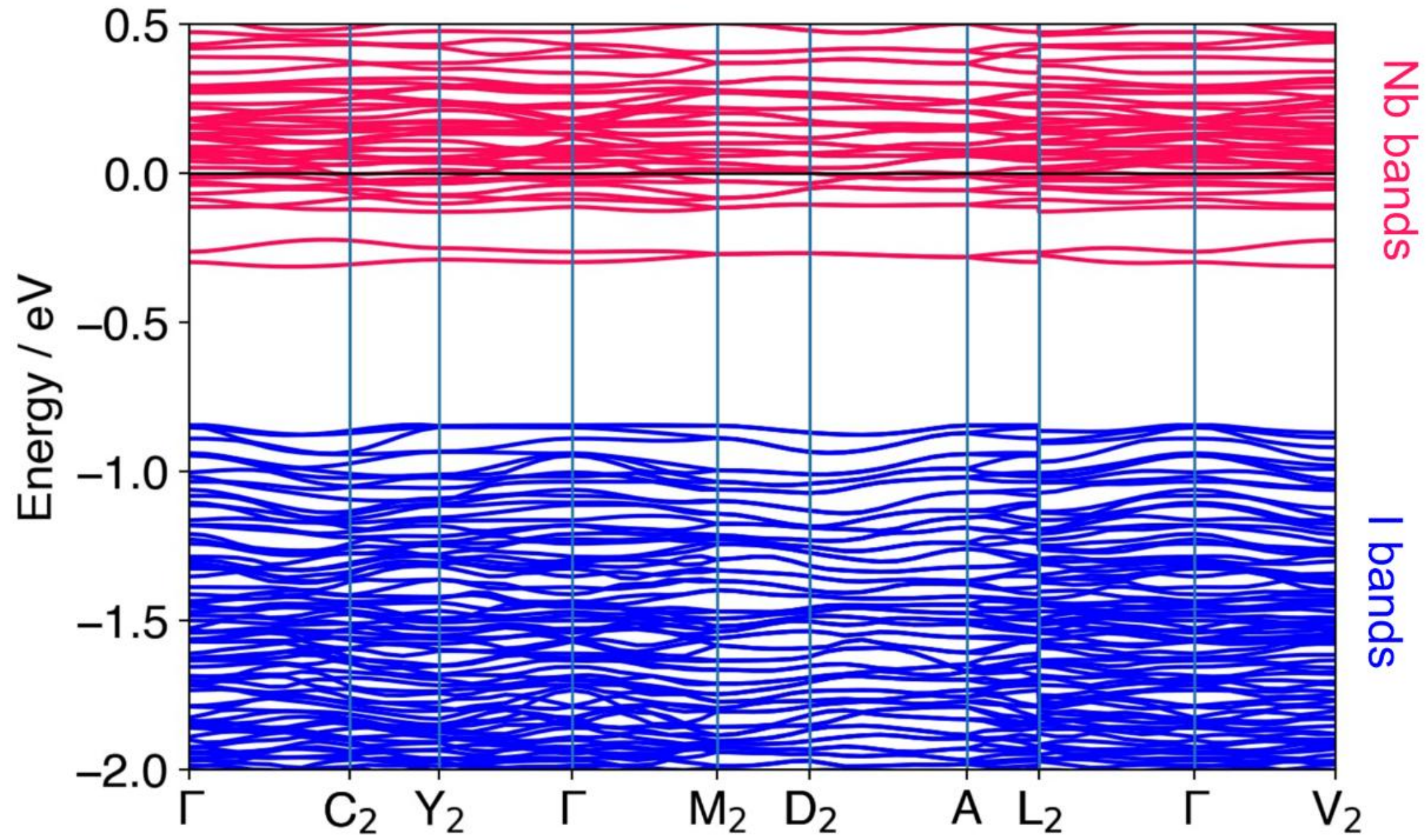

Fig. S2: Electronic band structure of $Nb_6O_3I_{15}$, with bands colored according to their main Nb or I character.

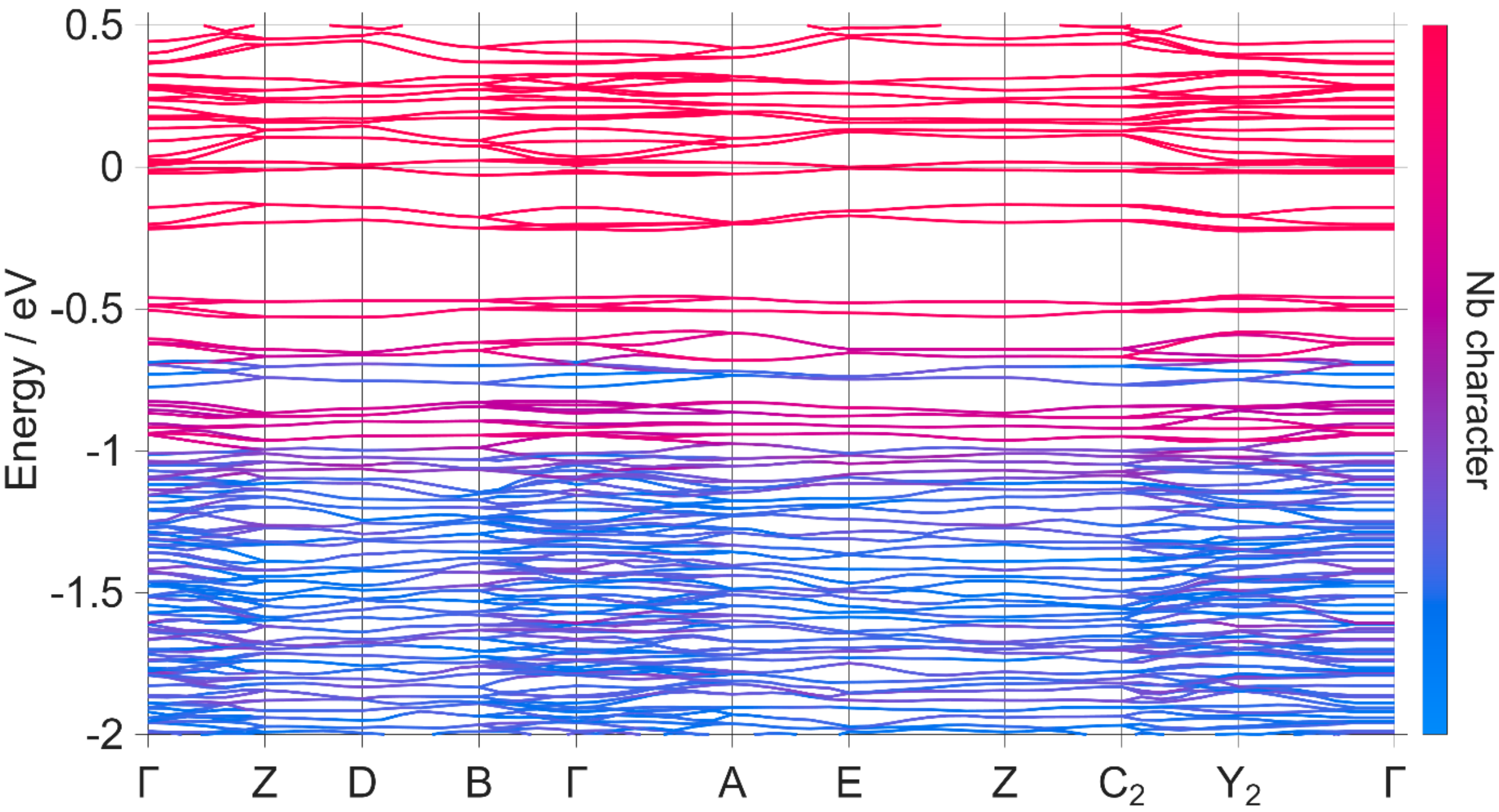

Fig. S3: Electronic band structure of $Nb_{11}O_6I_{24}$, with bands colored according to their Nb character.

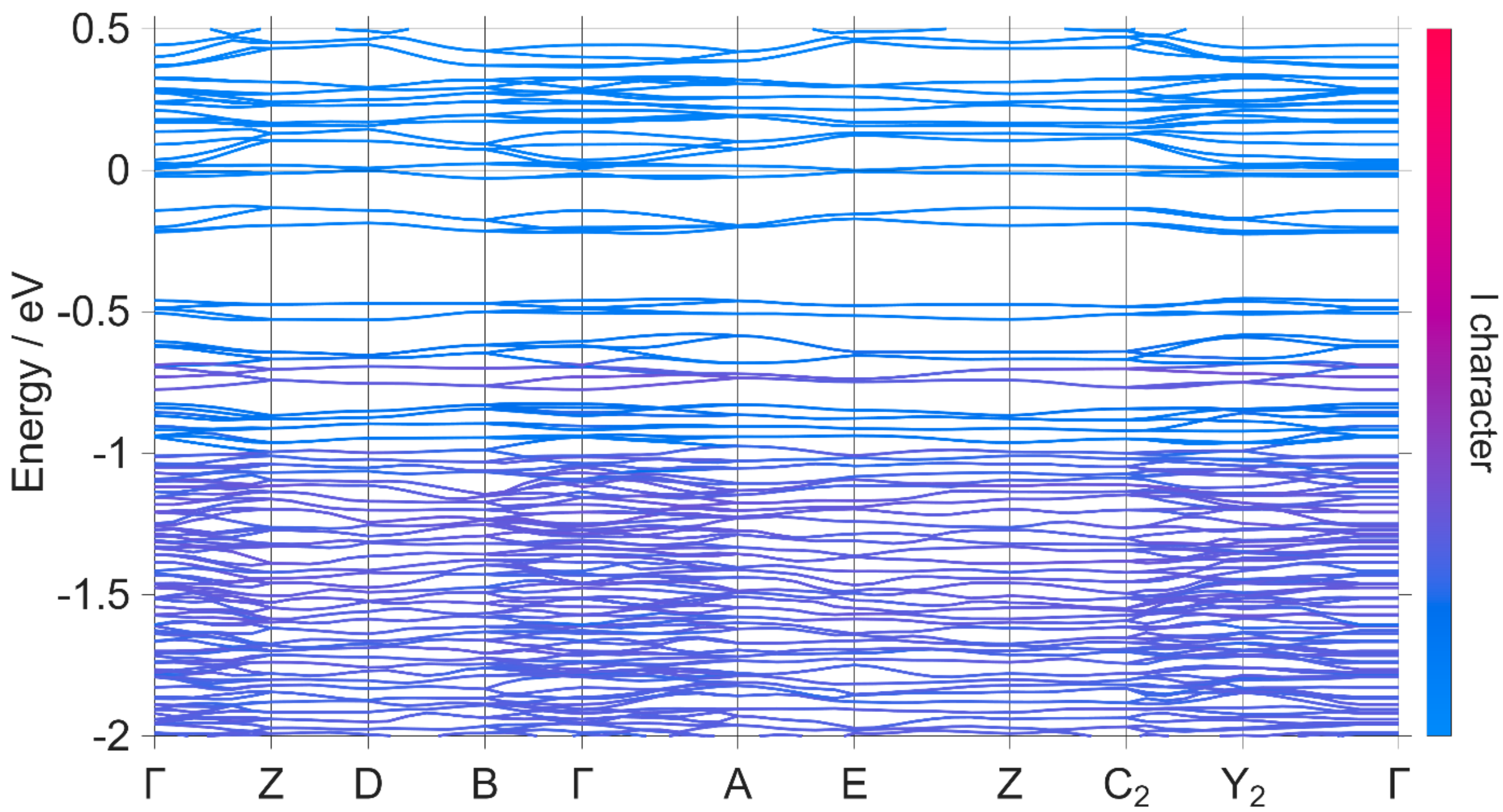

Fig. S4: Electronic band structure of $Nb_{11}O_6I_{24}$, with bands colored according to their I character.

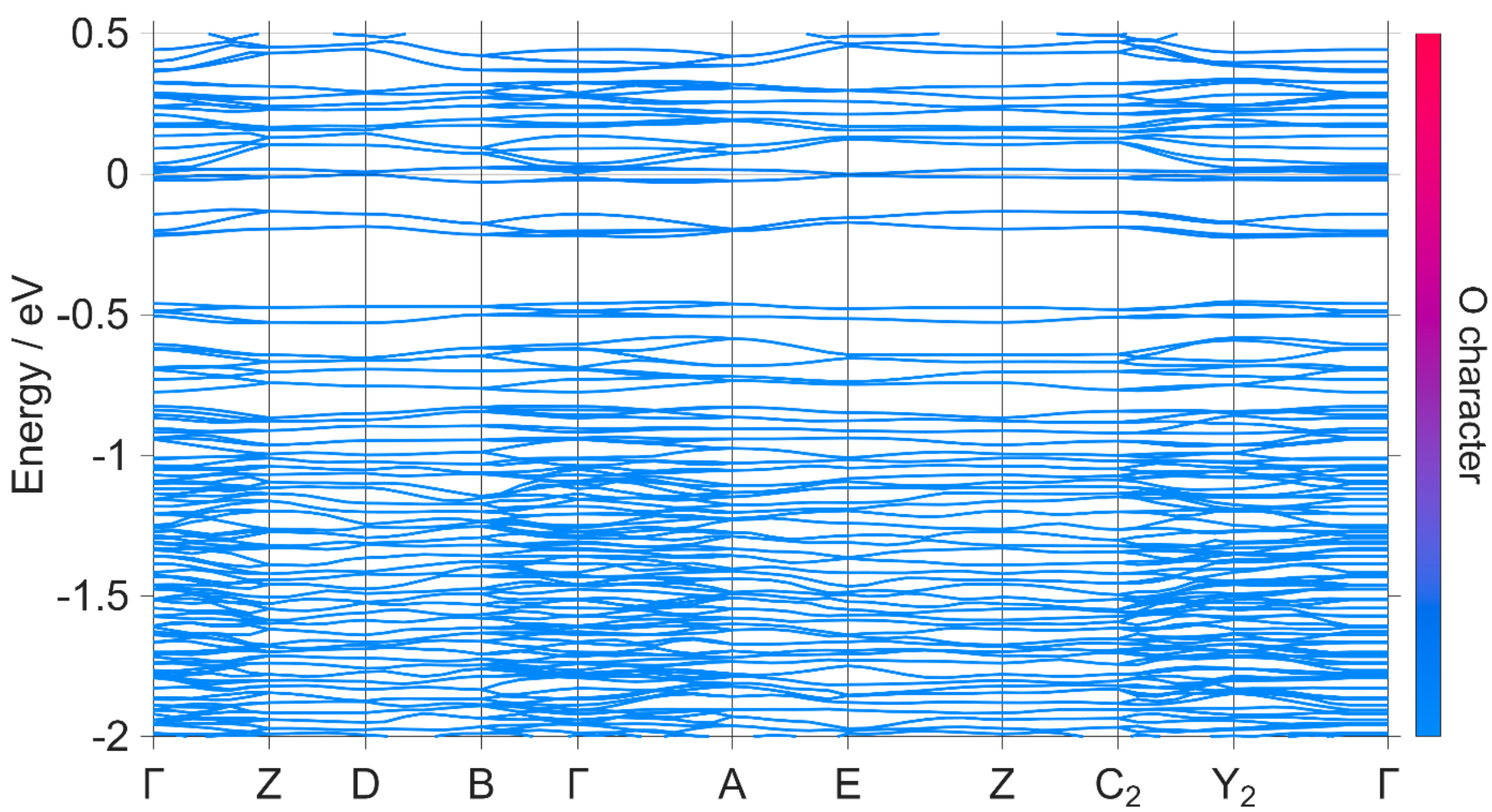

Fig. S5: Electronic band structure of $Nb_{11}O_6I_{24}$, with bands colored according to their O character.

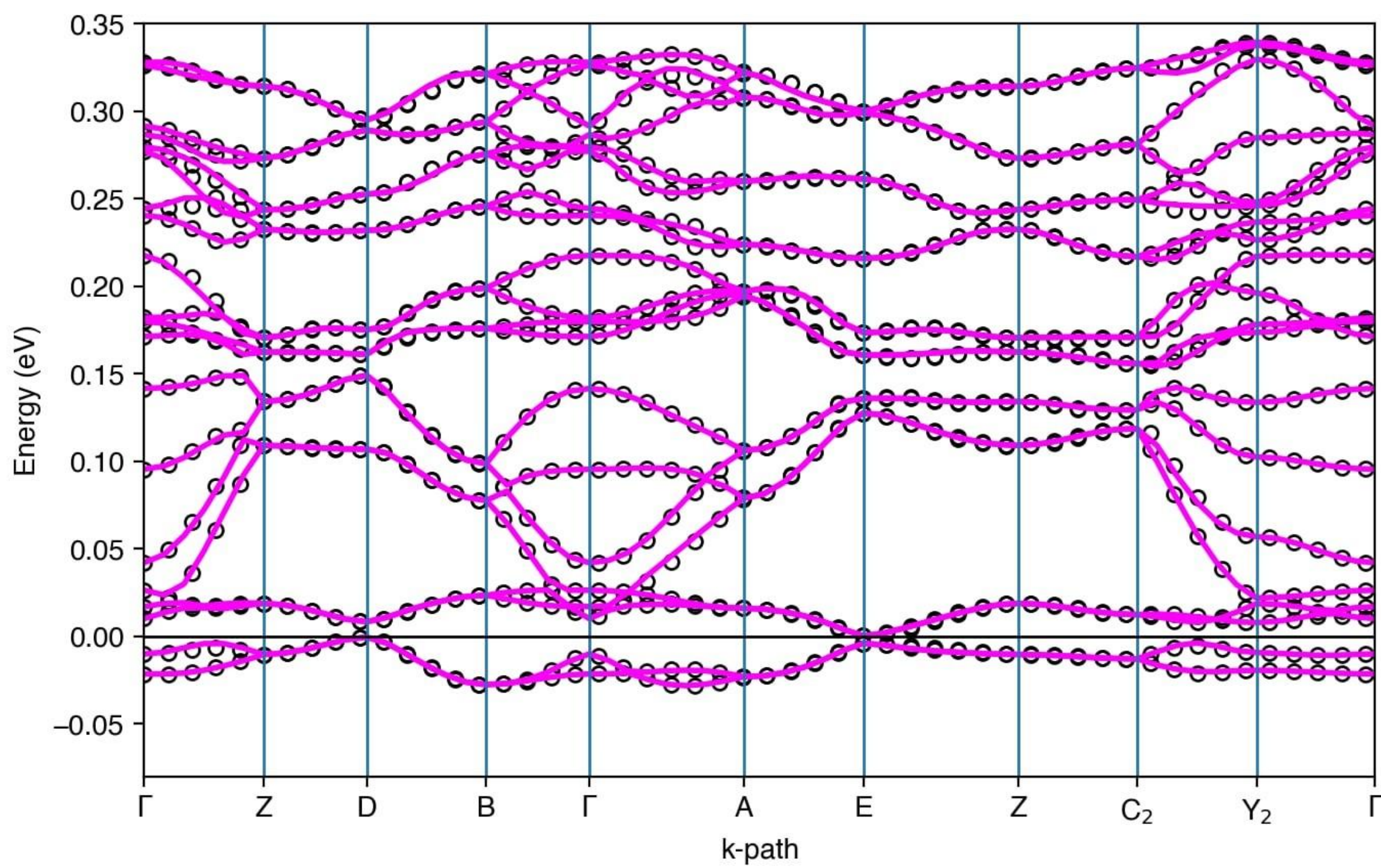

Fig. S6: Electronic band structure of $Nb_{11}O_6I_{24}$, with magenta lines showing the DFT-calculated bands and open circles showing the bands from the tight-binding model corresponding to the Wannier functions.

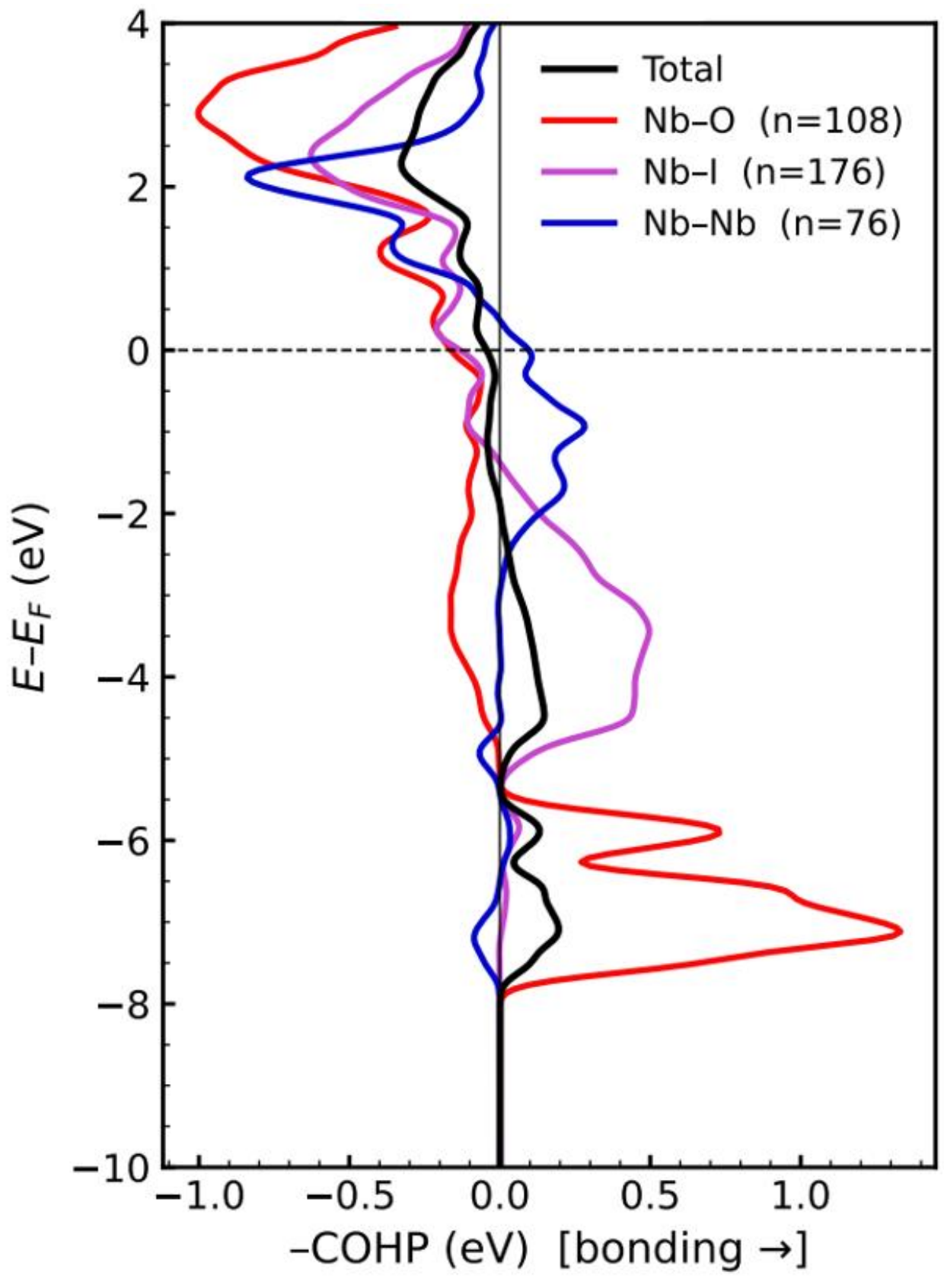

Fig. S7: Crystal orbital Hamilton populations (COHP) in $Nb_{11}O_6I_{24}$.

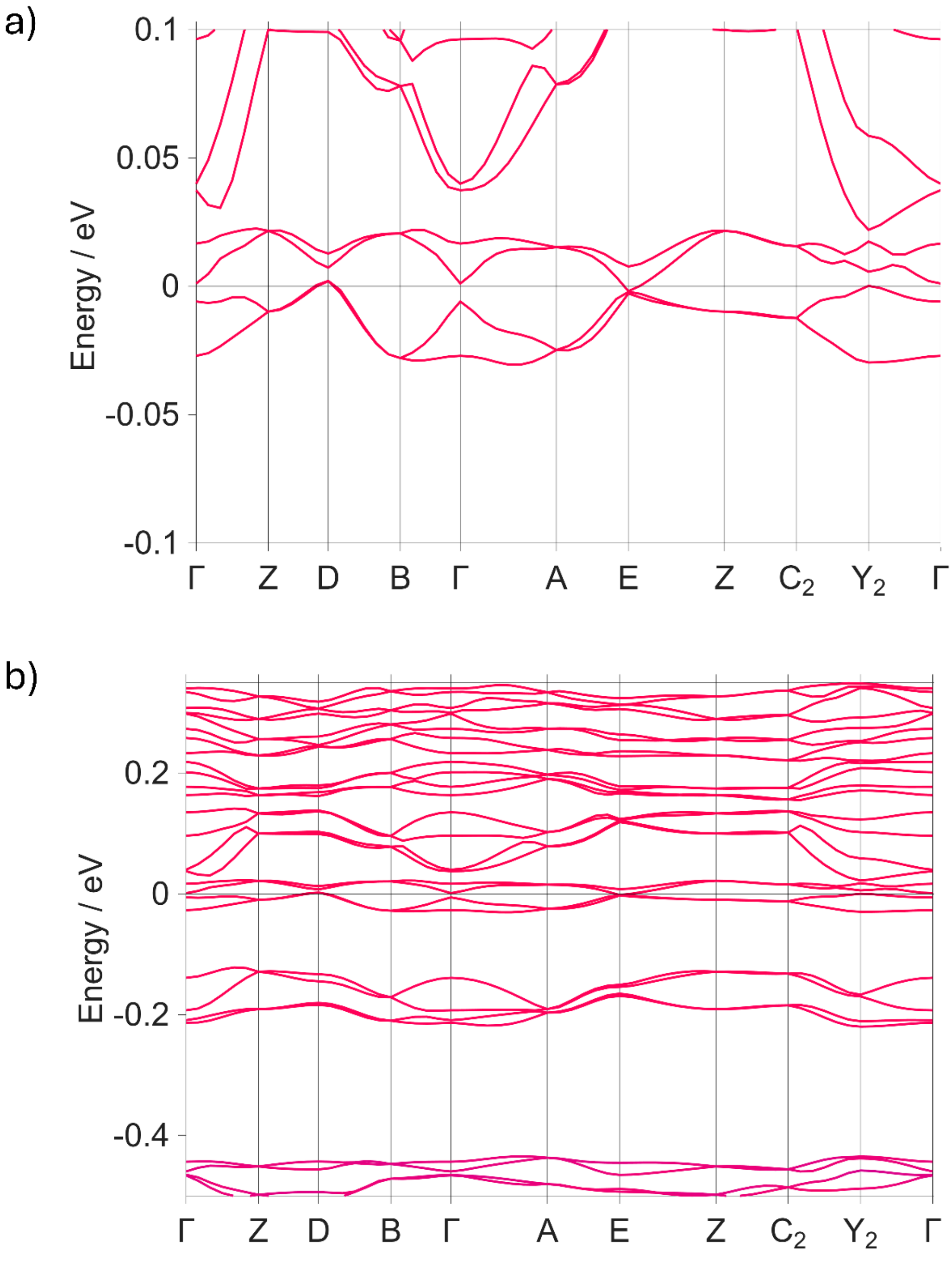

Fig. S8: Electronic band structure of $Nb_{11}O_6I_{24}$, calculated with spin–orbit coupling. A close-up of the bands near the Fermi energy is shown in (a), whereas a wider view is shown in (b).

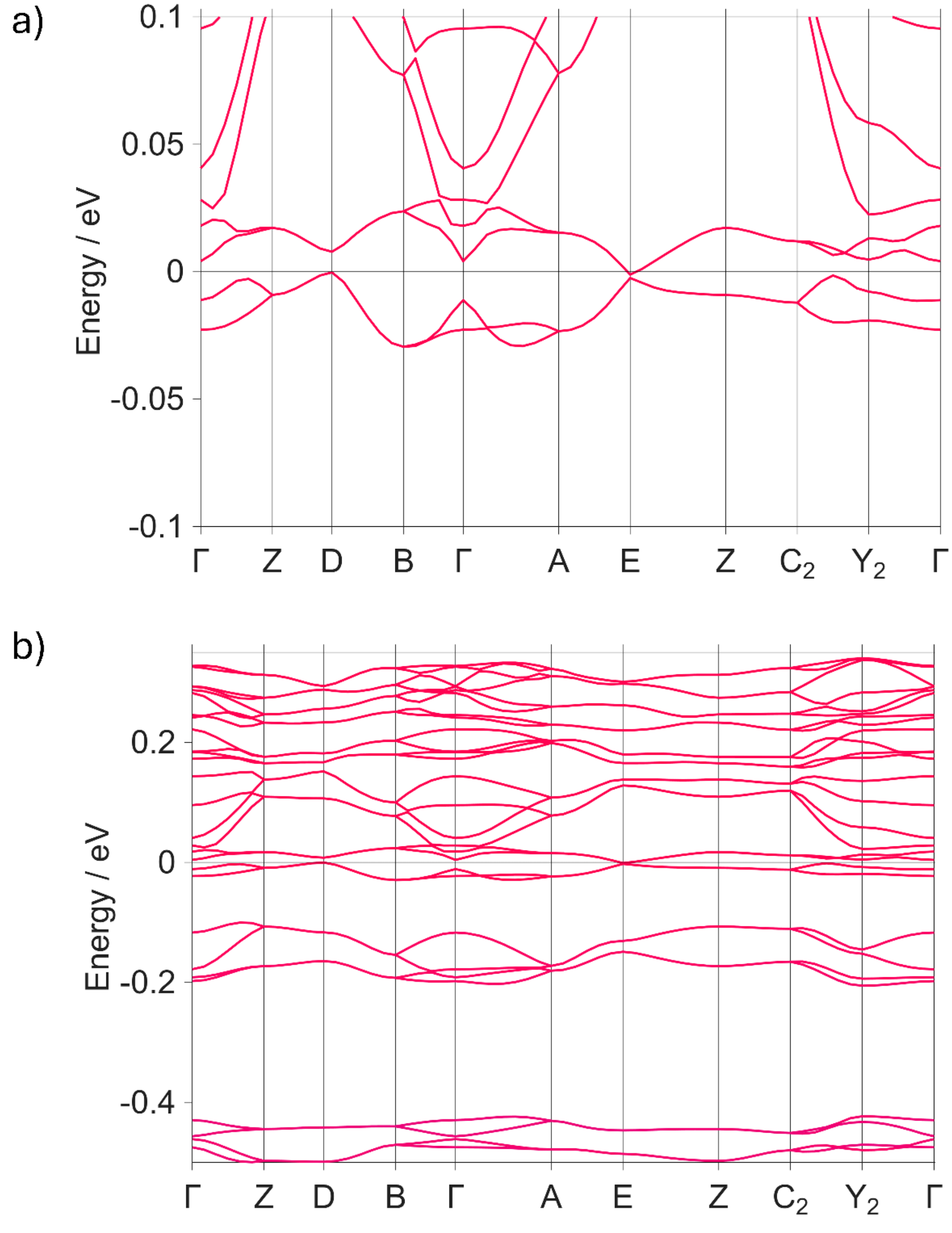

Fig. S9: Electronic band structure of $Nb_{11}O_6I_{24}$, calculated using the LDA exchange–correlation functional. A close-up of the bands near the Fermi energy is shown in (a), whereas a wider view is shown in (b).

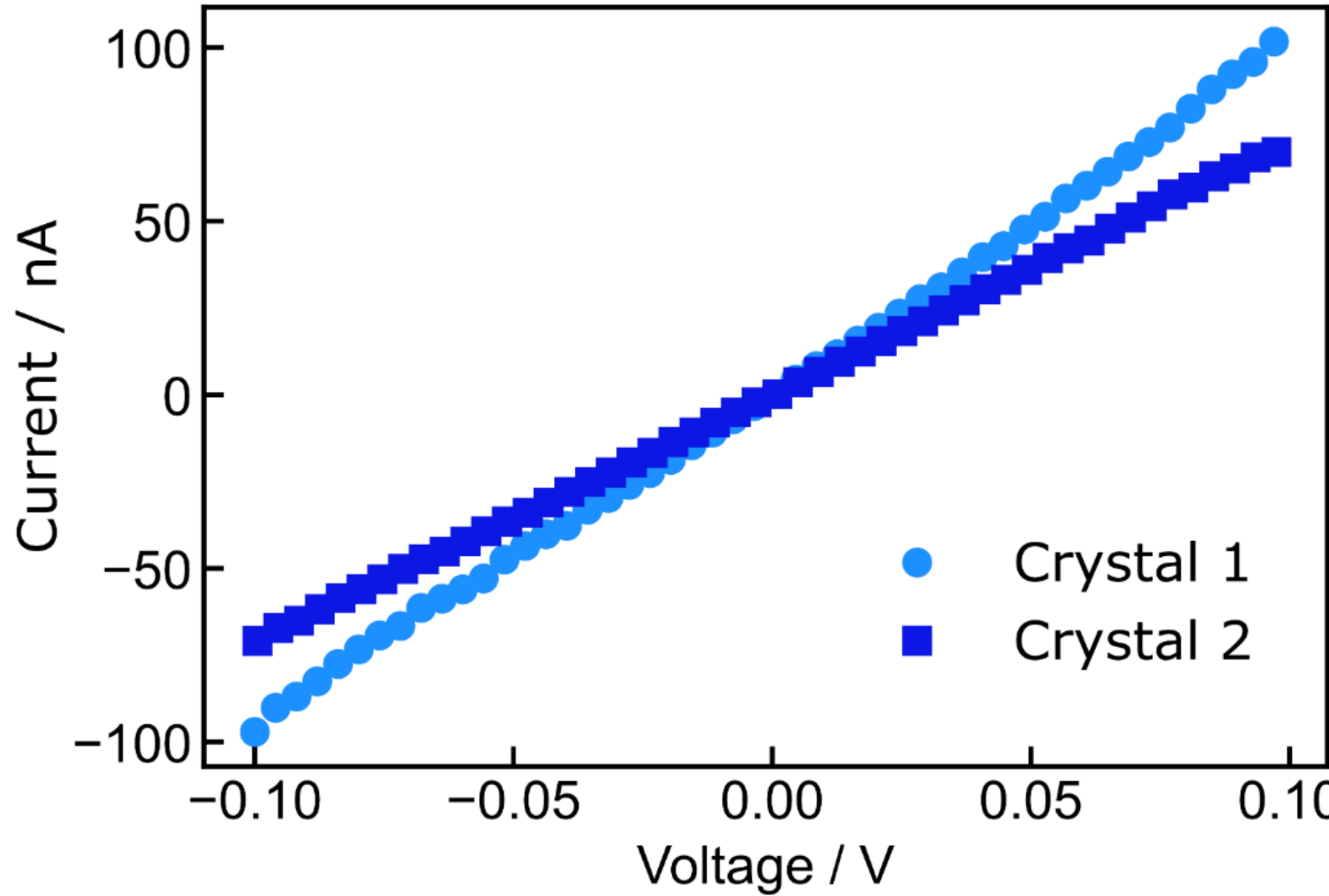

Fig. S10: *I-U* curves of two individual $Nb_{11}O_6I_{24}$-crystals at 300 K without illumination, showing Ohmic behavior.